\documentclass[useAMS,a4paper,fleqn,usenatbib]{mnras}
\usepackage{epsf}
\usepackage{graphicx}
\usepackage{multicol}
\usepackage{wrapfig}
\usepackage{natbib}
\usepackage{lscape}
\usepackage{hyperref}
\usepackage[T1]{fontenc}
\usepackage{aecompl}
\usepackage{gensymb}
\usepackage{amsmath}
\usepackage{amssymb}
\usepackage{longtable}

\newcounter{append}
\title[{\it INTEGRAL}/IBIS deep extragalactic survey: M81, LMC and 3C 273/Coma fields]{{\it INTEGRAL}/IBIS deep extragalactic survey: M81, LMC and 3C 273/Coma fields}
\author[I.A. Mereminskiy et al.]{Ilya A. Mereminskiy$^{1}$\thanks{E-mail:i.a.mereminskiy@gmail.com}, Roman A. Krivonos$^{1}$, Alexander A. Lutovinov$^{1}$, \newauthor
  Sergey Yu. Sazonov$^{1,2}$, Mikhail G. Revnivtsev$^{1}$ and Rashid A. Sunyaev$^{1,3}$\\
$^{1}$Space Research Institute, Russian Academy of Sciences, Profsoyuznaya 84/32, 117997 Moscow, Russia\\
$^{2}$Moscow Institute of Physics and Technology, Institutsky per. 9, 141700 Dolgoprudny, Russia\\
$^{3}$Max Planck Institute for Astrophysics, Karl-Schwarzschild-Strasse 1, D-85741 Garching, Germany}
\pubyear{2016}
\begin{document}
\pagerange{\pageref{firstpage}--\pageref{lastpage}}
\maketitle

\label{firstpage}

\begin{abstract}
We present results of deep surveys of three extragalactic fields, M81 (exposure of 9.7~Ms), LMC (6.8~Ms) and 3C 273/Coma (9.3~Ms), in the hard X-ray (17--60~keV) energy band with the IBIS telescope onboard the {\it INTEGRAL} observatory, based on 12 years of observations (2003--2015). The combined
survey reaches a $4\sigma$ peak sensitivity of 0.18~mCrab (2.6$\times$10$^{-12}$ erg s$^{-1}$ cm$^{-2}$) and sensitivity better than 0.25 and 0.87 mCrab over 10\% and 90\% of its full area of 4900 deg$^{2}$, respectively. We have detected in total 147 sources at $S/N>4\sigma$, including 37 sources observed in hard X-rays for the first time. The survey is dominated by extragalactic sources, mostly by active galactic nuclei (AGN). The sample of identified sources contains 98 AGN (including 64 Seyfert galaxies, 7 LINERs, 3 XBONGs, 16 blazars and 
8 AGN of unclear optical class), two galaxy clusters (Coma and Abell 3266), 17 objects located in the Large and Small Magellanic Clouds (13 high- and 2 low-mass X-ray binaries and 2 X-ray pulsars), 
three Galactic cataclysmic variables, one ultraluminous X-ray source (ULX, M82\,X-1) and one blended source (SWIFT J1105.7+5854). The nature of 25 sources remains unknown, so that 
the surveys identification is currently complete at 83\%. We have constructed AGN number-flux relations ($\log{N}$-$\log{S}$) and calculated AGN 
number densities in the local Universe for the entire survey and for each of the three extragalactic fields.
\end{abstract}


\section{Introduction}

Deep X-ray surveys of extragalactic fields with focusing X-ray telescopes \citep[see, e.g.,][for a review]{brandt15_rev} are essential for studying the evolution of active galactic nuclei (AGN) and physical processes powering their activity, but have a number of limitations. In particular, their small covered areas prevent finding a sufficient number of bright objects, whereas the soft X-ray energy band ($E\lesssim10$~keV) used in most surveys introduces a strong  bias against obscured (i.e. those with substantial intrinsic absorption) AGN. These drawbacks can be partially overcome using wide-field hard X-ray surveys performed with coded-mask telecopes like IBIS/{\it INTEGRAL} \citep{winkler03} or BAT/{\it Swift} \citep{gehrels04_swift}.

As was shown in previous studies \citep[see e.g.][]{paltani08,krivonos10_as1}, the IBIS telescope aboard the {\it INTEGRAL} observatory is able to achieve high sensitivity in
extragalactic fields. The sensitivity grows nearly proportionally to the square root of exposure showing no significant contribution of systematic noise and allowing IBIS to find sources at the
tenths-of-mCrab\footnote{One mCrab corresponds to $1.43\times10^{-11}$~erg~s$^{-1}$~cm$^{-2}$ in the 17--60 keV energy band assuming a spectral shape $10(E/1keV)^{-2.1}$ 
photons~cm$^{-2}$~s$^{-1}$~keV$^{-1}$.} flux level with a low number of false detections. In combination with IBIS large field of view 
(FOV, 28$\degree \times$ 28$\degree$, 9$\degree \times$ 9$\degree$ fully coded), this opens up a possibility to collect a significantly large sample of hard X-ray 
emitting AGN with fluxes down to a few $10^{-12}$~erg\,s$^{-1}$ cm$^{-2}$. Note that such objects, due to their rarity 
($\sim$0.05 AGN per deg$^{2}$), evade {\it NuSTAR} deep surveys \citep{mullaney15}.

The observational program of {\it INTEGRAL} has been mainly dedicated to Galactic source studies \citep[see, e.g.,][]{barlow06,revnivtsev08,bodaghee12,lutovinov13,walter15}, 
whereas the high Galactic latitude sky has been observed less intensively and very inhomogeneously. Nevertheless, on-going extragalactic surveys carried out with IBIS expand our knowledge about 
populations of extragalactic hard X-ray sources, mainly AGN, \citep{krivonos07_allsky, krivonos10_as2, bird09, bird16} and provide observational input for AGN studies
 \citep{sazonov07_agn,sazonov08_agnspe, sazonov15_obsagn, beckmann09, malizia09}.

A number of multi-year campaigns have been recently performed in the extragalactic sky, in particular of regions  around the M81 galaxy, the Coma cluster and the Large Magellanic Cloud. In each of these fields the total accumulated exposure (per position) exceeds 3 Ms, making them interesting for population studies of extragalatic hard X-ray sources and especially AGN in the so far poorly explored domain of sub-mCrab fluxes, which is the main purpose of the present paper.

The region around the M81 and M82 galaxies was targeted during two main campaigns: the study of hard X-ray spectra of the ultraluminous X-ray sources (ULXs) HoIX\,X-1 and M82\,X-1 \citep{sazonov14_ulx} and recent observations of the type Ia supernova SN\,2014J in M82 \citep{churazov14}. This field has a total exposure of 9.7~Ms (hereafter all quoted exposures are dead time corrected ones) at the position of the M81 galaxy.

\begin{figure}
\includegraphics[width=\columnwidth,bb=0 0 561 424]{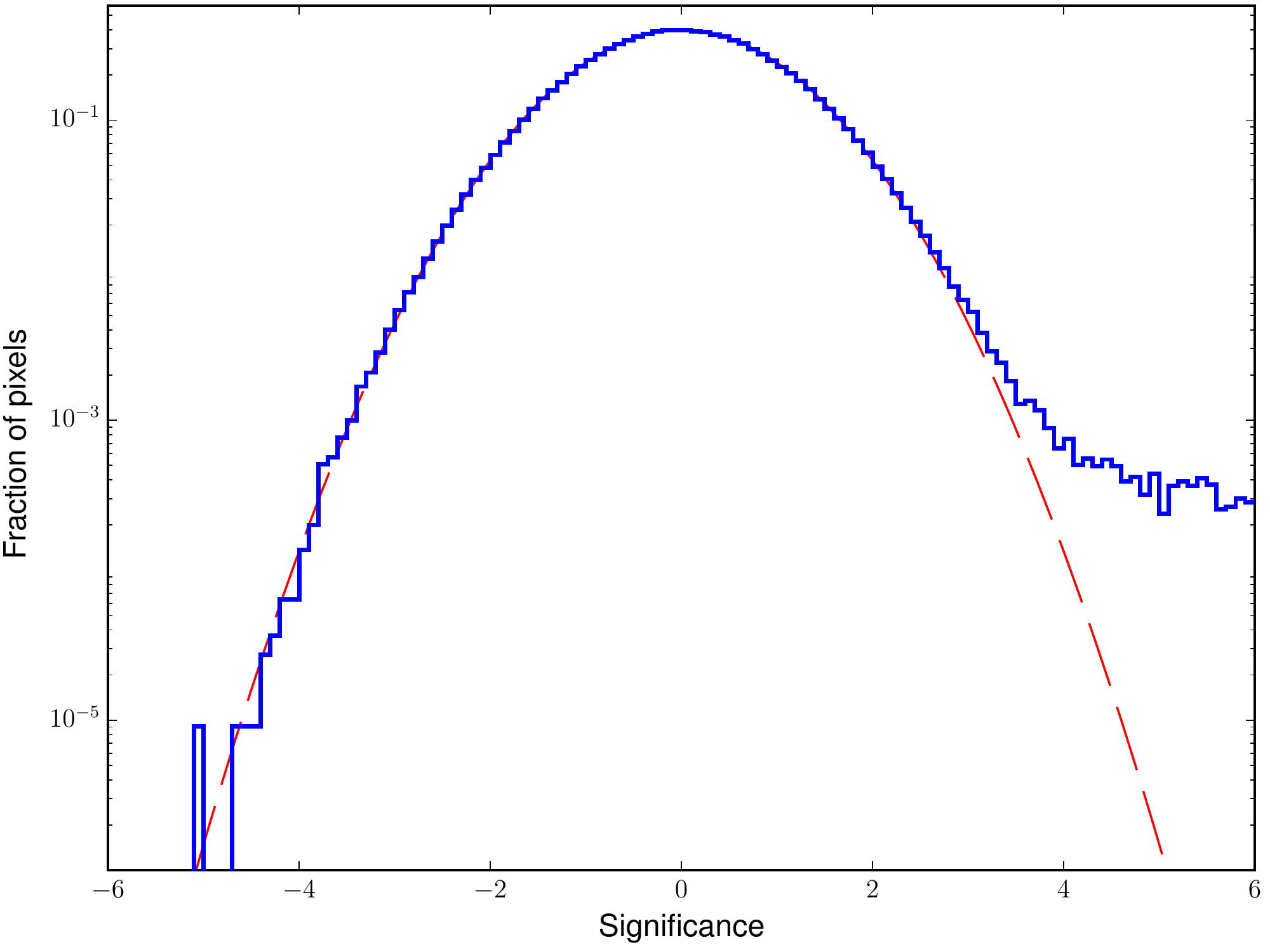}
\caption{Distribution of pixel significances for the combined survey of three extragalactic fields (blue histogram). The red dashed line shows the normal distribution with unit variance and zero mean.}\label{fig:signdist}
\end{figure}

Another {\it INTEGRAL} deep field, around the LMC galaxy, has a peak exposure of 6.8~Ms. The major part of the observing time was gained by the SN\,1987A
multi-year observational campaign \citep{grebenev_12sn}. The previous hard X-ray survey of the LMC region was presented by \citep{grebenev13LMC} and had reached a peak exposure of 4.8 Ms. Note that this field is rich in X-ray binaries located in LMC/SMC.

The field around the North Galactic pole was often observed as it includes a number of interesting extragalactic sources, such as the bright AGN 3C\,273 and NGC\,4151 and the Coma cluster. The region of the Coma cluster was first surveyed at hard X-rays with {\it INTEGRAL} by \citet{krivonos05_coma} who studied serendipitous extragalactic source counts down to a limiting flux of 1~mCrab. Later estimations \citep{krivonos07_allsky} showed that the Coma region has an enhanced population of AGN, which probably reflects the local overdensity of AGN in the nearby Universe. This result was later confirmed by {\it Swift}/BAT \citep{ajello12_bat60}. The sky region around 3C\,273/Coma (total area 2500 deg$^2$, exposure 4~Ms) was also selected to conduct an {\it INTEGRAL} extragalactic survey and to measure the source counts and AGN luminosity function \citep{paltani08}.

Given the IBIS FOV size and $5\times5$ standard observational pattern, we chose for our present study 35$\degree\times$35$\degree$ regions for the M81 and LMC fields with centers at J2000 coordinates RA=$85\degree.0$, Dec=$-69\degree.0$ and RA=$148\degree.9$, Dec=$69\degree.1$, respectively. For the  3C 273/Coma field, we chose an extended 35$\degree\times$70$\degree$ region with the aimpoint at RA=$190\degree.0$, Dec=$17\degree.0$ (J2000).

\begin{figure}
 \includegraphics[width=\linewidth,bb=0 0 566 430]{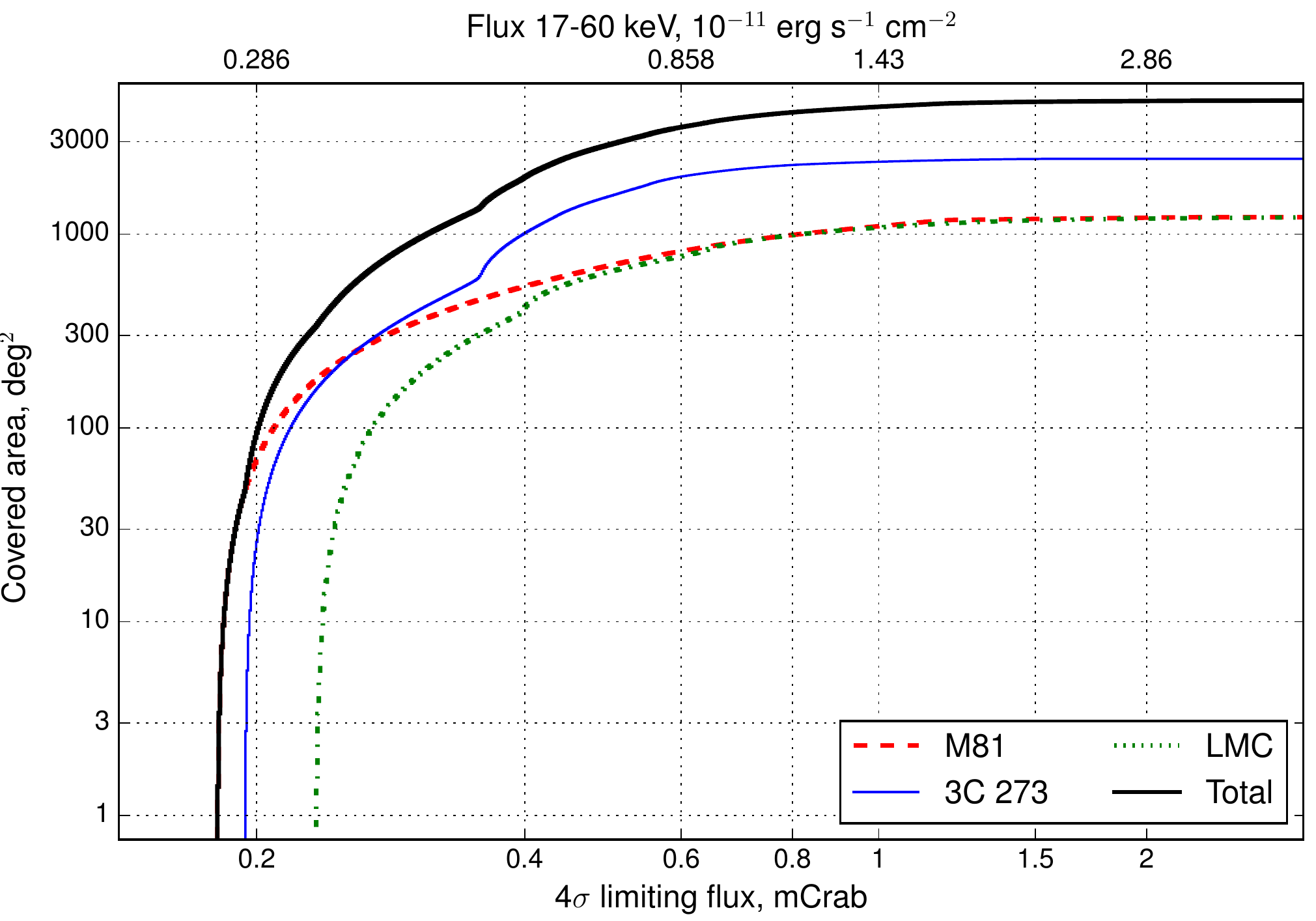}
 \caption{Sky area covered as a function of 4$\sigma$ limiting flux for each field and for the combined survey.}
 \label{fig:coverage}
\end{figure}
\section{Survey}

For the current survey we used all publicly available data acquired with {\it INTEGRAL} before June 2015 (spacecraft revolution 1553). The data from ISGRI, the first detector layer of the IBIS telescope, were utilized, as having the highest sensitivity at hard X-rays. We selected 17--60~keV as our working energy band where ISGRI has the highest effective area.

To apply the latest ISGRI energy calibration \citep{caballero13}, we first reduced the list of registered events with the {\it INTEGRAL} Offline Scientific Analysis (OSA) 10.1 provided by ISDC\footnote{ISDC Data Centre for Astrophysics, http://www.isdc.unige.ch/} up to the {\sc COR} level. Then we processed the events with a proprietary analysis package developed at IKI\footnote{Space Research Institute of the Russian Academy of Sciences} (details available in \citealt{krivonos10_as1}, \citealt{krivonos12} and \citealt{churazov14}) to produce individual 17--60~keV sky images for each {\it INTEGRAL} science window ({\it ScW}), an observation with a typical duration of 2--3~ks. Finally, the list of {\it ScW} images was cleaned to remove noisy ones and then combined into sky mosaics.

\begin{figure*}
\begin{minipage}[t]{\textwidth}
 \centering
 \includegraphics[width=0.98\linewidth]{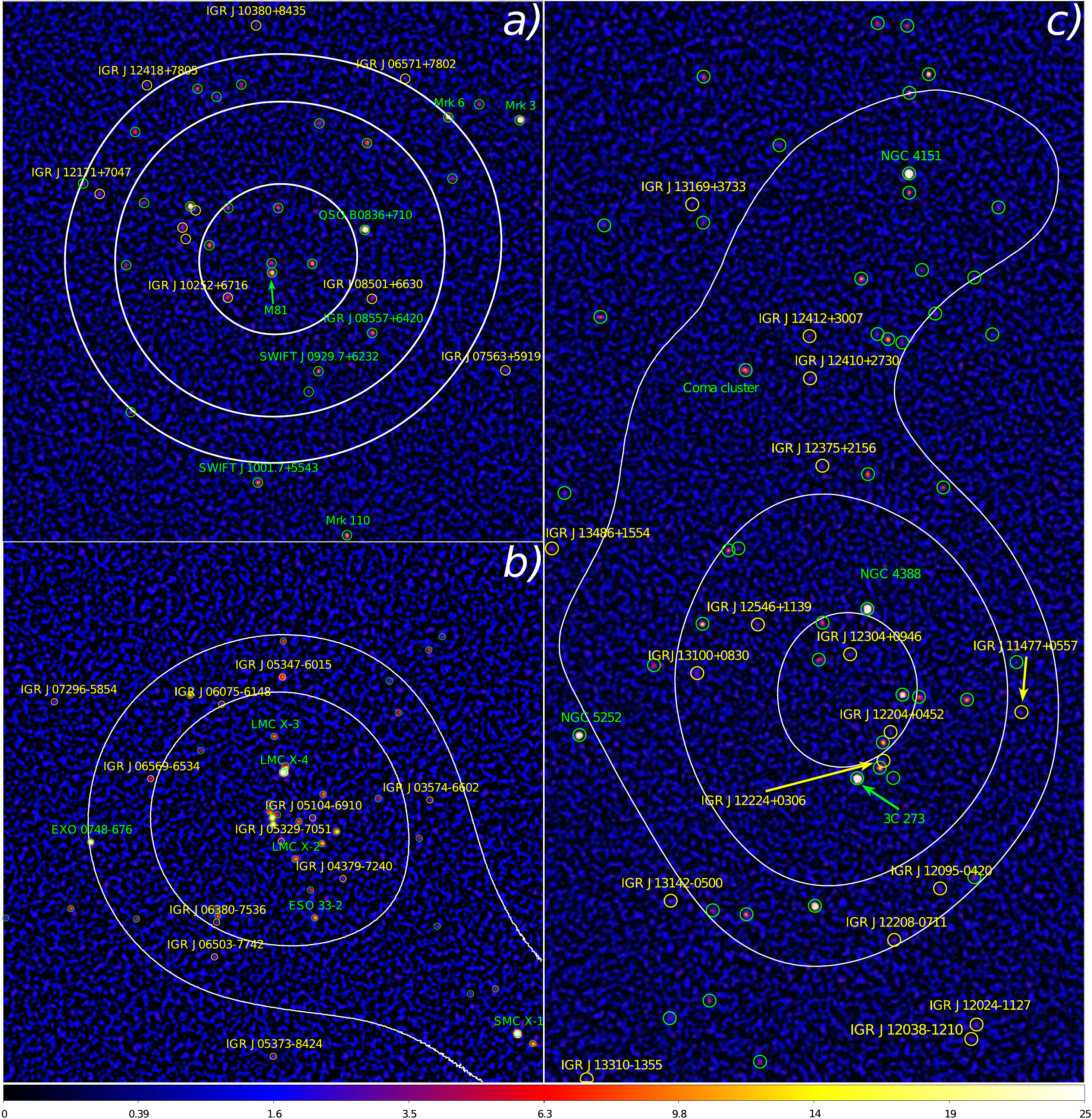}
 \caption{Hard X-ray maps of the M81, LMC and 3C 273/Coma fields, shown in terms of significance. The square-root color map ranges from 0 to 25. Yellow circles denote new sources and green circles already known ones. Some of the brightest sources are marked for easy navigation. North is up and east is to the left on all maps. {\it a) M81 field.} The peak exposure 9.7 Ms, contours show exposures of 2, 4 and 8 Ms. {\it b) LMC field.} The peak exposure 6.8 Ms, contours  drawn at 2 and 4 Ms. {\it c) 3C 273/Coma field.} The peak exposure 9.3 Ms, contours drawn at 2, 4 and 8 Ms.}
\label{fig:skymap}
\end{minipage}
\end{figure*}	

\subsection{Sensitivity and source detection}

The sensitivity of the current survey is limited by photon statistics and not significantly affected by systematic noise. The distribution of S/N values for  
pixels from all three sky mosaic maps is shown in Fig.\,\ref{fig:signdist}. It can be well described by pure statistical noise (at negative fluxes), which allows us to use Poisson statistics to predict the number of noise excesses above a given threshold. The combined survey covers a geometrical area of 4900~deg$^2$ which contains $\sim$1.2$\times$10$^{5}$ independent pixels of the size of the IBIS angular resolution (12\arcmin). By setting a 4$\sigma$ detection threshold for the current survey, we expect no more than 4 false detections in all three fields.

\begin{figure*}
 \includegraphics[width=\textwidth,bb=0 14 1132 522]{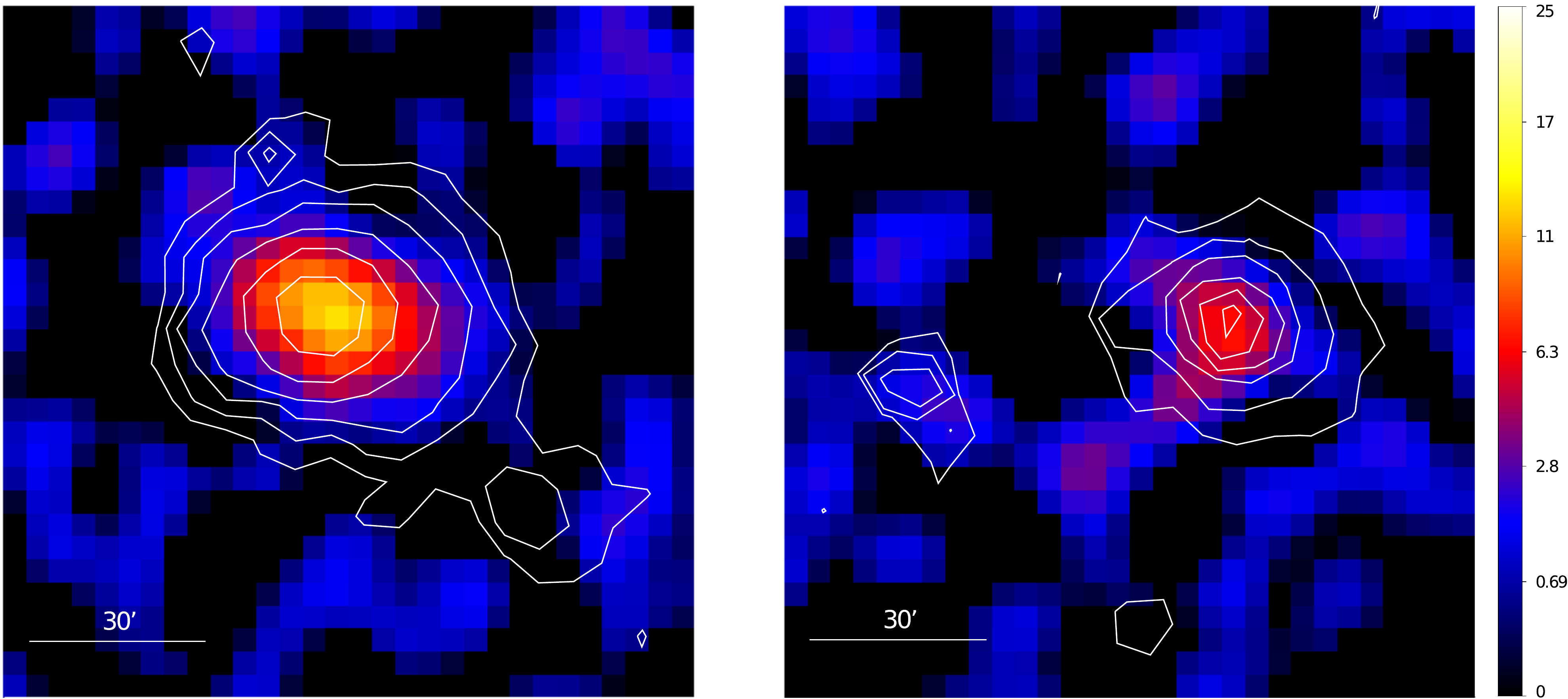}
 \caption{The 2$\degree \times$ 2$\degree$ fields around the Coma (left) and Abell\,3266 (right) clusters of galaxies. The contours denoting the surface brightness in the 0.1--2.4~keV energy band from {\it ROSAT} data are overplotted.}
\label{fig:clusters}
\end{figure*}
\begin{figure}
 \includegraphics[width=0.99\columnwidth,bb=0 0 572 433]{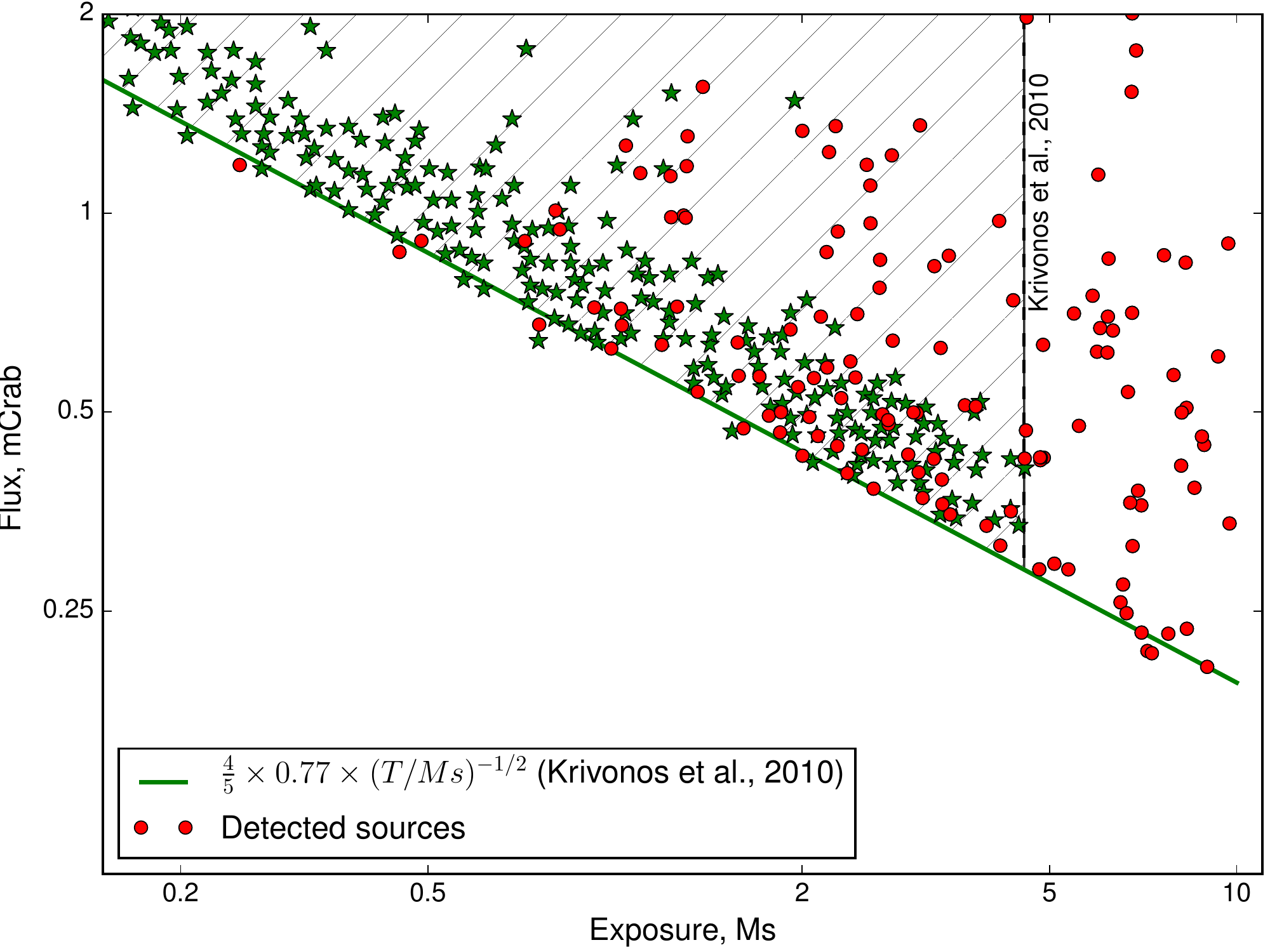}
 \caption{The 4$\sigma$ limiting flux as a function of the exposure. Red circles denote sources from the current survey. Green stars are high Galactic latitude ($|b|>15^{\circ}$) sources detected in the 7-year all-sky survey \citep{krivonos10_as2}. The green line represents an analytical approximation of the nominal sensitivity versus time.}
 \label{fig:detlim}
\end{figure}

Fig.\,\ref{fig:coverage} shows the area covered by the combined survey and individual fields as a function of the 4$\sigma$ limiting flux. The peak sensitivity of the survey is 0.18 mCrab (2.6$\times$10$^{-12}$ erg s$^{-1}$ cm$^{-2}$), with 10\% and 90\% of the total area having been covered with sensitivity better than 0.25 mCrab (3.6$\times$10$^{-12}$ erg s$^{-1}$ cm$^{-2}$) and 0.87 mCrab (1.2$\times$10$^{-11}$ erg s$^{-1}$ cm$^{-2}$), respectively.

We analyzed the mosaic maps for positive excesses with $S/N>4\sigma$ and found 147 source candidates. We cross-checked the list of the detected sources with the current {\it INTEGRAL} source catalog\footnote{\url{http://isdc.unige.ch/integral/catalog/39/catalog.html}}, the {\it Swift}/BAT 70-month catalog \citep{baumgartner13_swift70} and the all-sky hard X-ray survey by \citet{bird16} as the most complete and up-to-date hard X-ray source catalogs. We also used the 66-month Palermo {\it Swift}/BAT online catalog\footnote{\url{http://bat.ifc.inaf.it/bat_catalog_web/66m_bat_catalog.html}} \citep{cusumano10} as a complementary catalog. For all identified extragalactic sources we collected known redshifts or distances from the NASA/IPAC Extragalactic Database\footnote{\url{https://ned.ipac.caltech.edu/}} (NED).

Fig.\,\ref{fig:skymap} shows the mosaic images along with exposure contours for the three studied fields. Note that the M81 and 3C 273/Coma fields do not show any systematic noise, which suggests that IBIS/ISGRI can be used to perform even deeper extragalactic surveys. The source statistics for each field is discussed below.

The high Galactic latitude ($b^{II}\approx40^{\circ}$) M81 field contains 37 detected sources: 28 known AGN including 5 blazars, one ULX (M82\,X-1), two Galactic binary systems (MU Cam and DO Dra), five new hard X-ray sources of unknown type, and SWIFT\,J1105.7+5854 -- a known pair of sources with 6' separation \citep{baumgartner13_swift70}, which cannot be resolved with the IBIS telescope. This field hosts the most distant object in our survey -- the 
quasar QSO\,B0836+710 at $z=2.172$ \citep{stickel93}. Thus, the M81 field is dominated by AGN.
	
In the LMC field, 46 sources are detected, including 11 objects previously unknown as hard X-ray sources. This field is different from the other two because it hosts 17 objects located in the Magellanic Clouds (both LMC and SMC), including 13 high-mass X-ray binaries, two low-mass X-ray binaries and two rotation-powered pulsars; it also contains the Galactic cataclysmic variable TW Pic. Among 21 extragalactic sources in this field there are 20 AGN, including three blazars, and the cluster of galaxies Abell 3266. The nature of 7 sources remains unknown.

In the vicinity of the bright ($\sim$20~mCrab) X-ray pulsar LMC\,X-4, the presence of two hard X-ray sources was reported earlier: IGR\,J05319$-$6601 \citep{gotz06} and IGR\,J05305$-$6559 \citep{krivonos07_allsky}. Due to the small angular distance from LMC\,X-4, these sources cannot be resolved on the average map. Nevertheless, taking into account a peculiar property of LMC\,X-4 (the source periodically goes to the ''off''-state) and using the corresponding subset of revolutions, \cite{grebenev13LMC} showed that the persistent hard X-ray emission actually originates from the sky region coinciding 
with the position of another X-ray pulsar, EXO\,053109-6609.2. This conclusion was supported by an independent detection of this source in the standard X-ray energy band by the {\it INTEGRAL}/JEM-X telescope, which allowed \citet{grebenev13LMC} to reconstruct the source spectrum in a  broad energy band and demonstrate that it is typical for accreting X-ray pulsars. Based on the extended data set obtained with {\it INTEGRAL} and using the current ephemerides for LMC\,X-4 \citep{molkov15}, we have repeated such an analysis and verified the result of \citet{grebenev13LMC}. Summarizing the above, we can conclude that the hard X-ray emission detected by {\it INTEGRAL} from the vicinity of LMC\,X-4 is associated with the X-ray pulsar EXO\,053109$-$6609.2 and that the hard X-ray sources IGR\,J05319$-$6601 and IGR\,J05305$-$6559 are actually the same source -- the X-ray counterpart of EXO\,053109$-$6609.2. We finally note that since our maps are averaged over many revolutions, the position and flux for this source in the catalog (referred to as IGR\,J05305$-$6559) are strongly affected by LMC\,X-4 and should thus be treated carefully.

The 3C\,273/Coma field is the largest one and naturally contains the largest number of sources. We have detected here 64 sources including 16 objects  detected in hard X-rays for the first time. All the identified sources are of extragalactic origin: there are 47 known AGN, including 7 blazars, and the Coma cluster. This field hosts the faintest object in our survey -- IGR\,J12304+0946 with the flux of $0.21\pm0.05$~mCrab (3.0$\times$10$^{-12}$ erg s$^{-1}$ cm$^{-2}$).

{\it INTEGRAL} has detected for the first time the galaxy cluster Abell\,3266, with the flux of $0.64\pm0.10$~mCrab, confirming the previous 
detection by {\it Swift}/BAT \citep{ajello09_cl}. However, unlike the Coma cluster \citep{lutovinov08_coma}, the extended structure of Abell\,3266 is not resolved by {\it INTEGRAL}. In Fig.\,\ref{fig:clusters}, we present zoomed IBIS images of these two clusters with oveplotted contours demonstrating the 0.1--2.4 keV surface brightness, obtained from {\it ROSAT} data \citep{voges99}. Regions with the highest hard X-ray brightness coincide with the central parts of the soft X-ray images.

In summary, we have detected 147 sources in all three fields, which are listed in Table~\ref{tab:srcstab_red}, including 37 detected in hard X-rays for the first time. Two fields (M81 and 3C\,273/Coma) are dominated by extragalactic objects, while a significant fraction of sources in the LMC field are nearby ones (X-ray binary systems) located in LMC and SMC.

Fig.\,\ref{fig:detlim} shows the fluxes of the detected sources as a function of the exposure time, along with an expected sensitivity curve $F_{{\rm lim}}^{5\sigma} = 0.77 \times (T /{\rm Ms})^{-0.5}$~mCrab provided by \cite{krivonos10_as2}. We see that the IBIS/ISGRI extragalactic survey continues to operate in a statistically limited regime, with the sensitivity increasing as the square root of the exposure. The factor of $\sim2$ improvement in sensitivity with respect to the 7-year all-sky survey \citep{krivonos10_as2} is clearly visible.

It is interesting to compare our catalog with an {\it INTEGRAL} all-sky survey catalog recently published by \citet{bird16} based on IBIS data taken before spacecraft orbit 1000 (December 2012). The 
catalog of \citet{bird16} contains only 65 sources out of the 147 sources detected in our survey (14/37 in the M81 field, 17/46 in LMC and 34/64 in 3C 273/Coma), 
which is not unexpected given that several extensive {\it INTEGRAL} observational campaigns of these fields have been undertaken after December 2012 and we have taken advantage 
of these additional data. On the other hand, since our survey was not designed for source detection at different time scales, it misses 12 short and 3 long transients listed in \citet{bird16} with 
typical outburst timescales of weeks and months, respectively. In addition, 10 persistent weak sources from \citet{bird16} catalog fall below our detection threshold, including three sources 
in the M81 field (IGR J08447+6610, Mrk 18 and IGR J09034+5329), two in LMC (PKS 0312-770 and SWIFT J0450.7-5813) and five in 3C 273/Coma (IGR J12562+2554, IGR J13166+2340, 
SWIFT J1344.7+1934, IGR J12319-0749 and IGR J11486-0505), which may indicate that these sources became dimmer in the latest {\it INTEGRAL} observations.

\begin{table*}
\begin{minipage}{\textwidth}
\small
\vspace{6mm}
\centering
\caption{Part of the catalog of sources detected in the combined survey of three fields: M81, LMC and 3C 273/Coma. The descripton of the columns can be found in Sect.~\ref{sect:catalog}. The full version of the table is available in \hyperref[append:fullcat]{{\bf Appendix 1}}.}
\label{tab:srcstab_red}
\vspace{1mm}
\begin{tabular}{clrrrcccccl} \hline
Id&Name$^1$&R.A.&Dec.&S/N&Flux & $D$ &$z$&$\log{L}$&Type&Notes\\
   & &deg&deg &        & mCrab &Mpc& & erg s$^{-1}$  & &\\
\hline
\multicolumn{11}{c}{{\bf M81 field}}\\
\hline
1&Mrk 3                              &93.950 &71.039 &39.4 & 5.81$\pm$0.15 &  &0.013&43.56&Sy2&\\
2&IGR J06253+7334             &96.370 &73.585 &7.6   &0.99$\pm$0.13 & &&&CV&MU Cam\\
3&Mrk 6                              &103.043 &74.427&22.4 &2.37$\pm$0.11 & &0.019&43.46&Sy2&\\
4&{\bf IGR J06571+7802}   &104.277 &78.044&4.2   &0.47$\pm$0.11 & & & & &\\
5&QSO B0716+714              &110.576 &71.304&5.9   &0.50$\pm$0.08 & &0.300&45.34&Blazar&\\
6&{\bf IGR J07563+5919}   &119.091 &59.321&4.0   &0.62$\pm$0.16 & & & & & \\
7&PG 0804+761                  &122.929  &76.034&9.4  &0.63$\pm$0.07 & &0.100&44.39&Sy1&\\
$<\dots>$&&&&&&&&&&\\
147&{\bf IGR J13486+1554}                 &207.168  &15.901&4.8  &0.57$\pm$0.12 & &&&&\\
\hline
\end{tabular}
\end{minipage}
\begin{flushleft}
$^{1}$ The names of sources previously unknown in the hard X-ray band (17--60~keV) are highlighted in bold. Sources with spatial confusion are indicated by a star, their measured fluxes should be considered with caution.\\
\end{flushleft}

\end{table*}
\begin{table*}
\caption{1SXPS sources in the $4.2\arcmin$ error circle around the IGR\,J13100+0830 position. The table is based on the 1SXPS catalog \citep{evans_14}.}
\label{tab:13100srcs}
\begin{minipage}[h]{\linewidth}
\centering
\begin{tabular}{ccccccc} \hline
1SXPS & Offset$^{1}$ &R.A., Dec.& \multicolumn{2}{c}{Count rate} &  Flux$^{3}$  & Optical counterpart\\
Id    &            &        (error$^{2}$)                               &                         \multicolumn{2}{c}{ $\times10^{-4}$ cts s$^{-1}$}                               & $\times10^{-14}$~erg~s$^{-1}$~cm$^{-2}$  & (type) \\
    &            &                                                         &    0.2--2~keV                     &                          2--10~keV     & 0.2--10~keV &  \\
\hline
J131004.4+082936  & 1.1$\arcmin$ &197.5184, 8.4935 & 13.0 & 5.5 & 7.5$^{+1.8}_{-1.6}$ &  SDSS J131004.26+082938.9\\
                                    &                        &(4.8$\arcsec$)&         &       &        & (QSO candidate, $z=1.22$) \\
J131008.5+082826  & 2.6$\arcmin$ &197.5356, 8.4741& 13.8 & 3.4 & 7.0$^{+1.7}_{-1.5}$  & SDSS J131008.34+082826.4\\
                                     &                        &(4.4$\arcsec$)&          &       &        & (galaxy, $z=0.27$)\\
J131014.2+083137  & 3.3$\arcmin$ &197.5592, 8.5270& 16.2 & 3.5 & 8.0$^{+1.8}_{-1.6}$ & SDSS J131014.24+083135.9\\
                                    &                        &(4.9$\arcsec$)&          &       &        & (QSO candidate, $z=1.55$) \\
J130947.2+083049  & 3.6$\arcmin$ &197.4467, 8.5138& 5.5 & 1.7 & 3.0$^{+1.1}_{-1.0}$ &USNO-B1.0 0985-0230131\\
                                    &                        &(5.3$\arcsec$)&          &       &        & (foreground star) \\
\hline
\end{tabular}
\end{minipage}
\begin{flushleft}
$^{1}$ Angular offset from the {\it INTEGRAL} position of IGR\,J13100+0830 in arcminutes.\\
$^{2}$ Radius of the 90\% confidence error circle.\\
$^{3}$ The 0.2--10~keV flux ($\pm1\sigma$ error) calculated from a power-law model with $\Gamma$=1.7 and Galactic absorption toward the source \citep[see details in][]{evans_14}.\\
\end{flushleft}
\end{table*}

\subsection{Identification of new sources}

For the identification of 37 newly detected sources we utilized the SIMBAD\footnote{\url{http://simbad.u-strasbg.fr/simbad/}} and HEASARC\footnote{\url{http://heasarc.gsfc.nasa.gov/}} databases as well as the {\it Swift}/XRT point source catalog \citep[1SXPS,][]{evans_14} and the
third {\it XMM-Newton} serendipitous source catalog \citep[3XMM-DR5,][]{rosen_15}. Based on {\it XMM-Newton} or {\it Swift}/XRT archival observations, we selected X-ray counterparts in the soft X-ray band (2--10~keV) within a $4.2'$ ($2\sigma$) error circle around the best-estimate positions of hard X-ray sources. The favored source was that which had the highest flux and a hard spectrum consistent with the {\it INTEGRAL} 17--60~keV flux. We found firm soft X-ray counterparts for 13 sources of 37, and list them in Table~\ref{tab:srcstab_red}. In some cases we propose an optical counterpart based on positional coincidence with a known bright source, e.g. an AGN. Below we discuss a few cases of source identifications in which additional observations are needed to validate the proposed association.

\section*{IGR\,J08501+6630}

Our search for a soft X-ray counterpart in the HEASARC archival data did not yield a potential candidate within the {\it INTEGRAL} error circle of IGR\,J08501+6630. However, we found two bright sources in optical/IR bands: the star TYC 4134-706-1 \citep{tycho_cat} and the edge-on spiral galaxy MCG+11-11-029 ($z=0.037$). The latter is proposed as a possible optical counterpart of IGR\,J08501+6630. The absence of a soft X-ray counterpart and non-detection of IGR\,J08501+6630 in the {\it ROSAT} all-sky survey \citep{voges99} indicates a strong intrinsic absorption.

\section*{IGR\,J05329$-$7051}

\begin{figure}
\centering
\includegraphics[width=0.81\linewidth,bb=17 21 465 500]{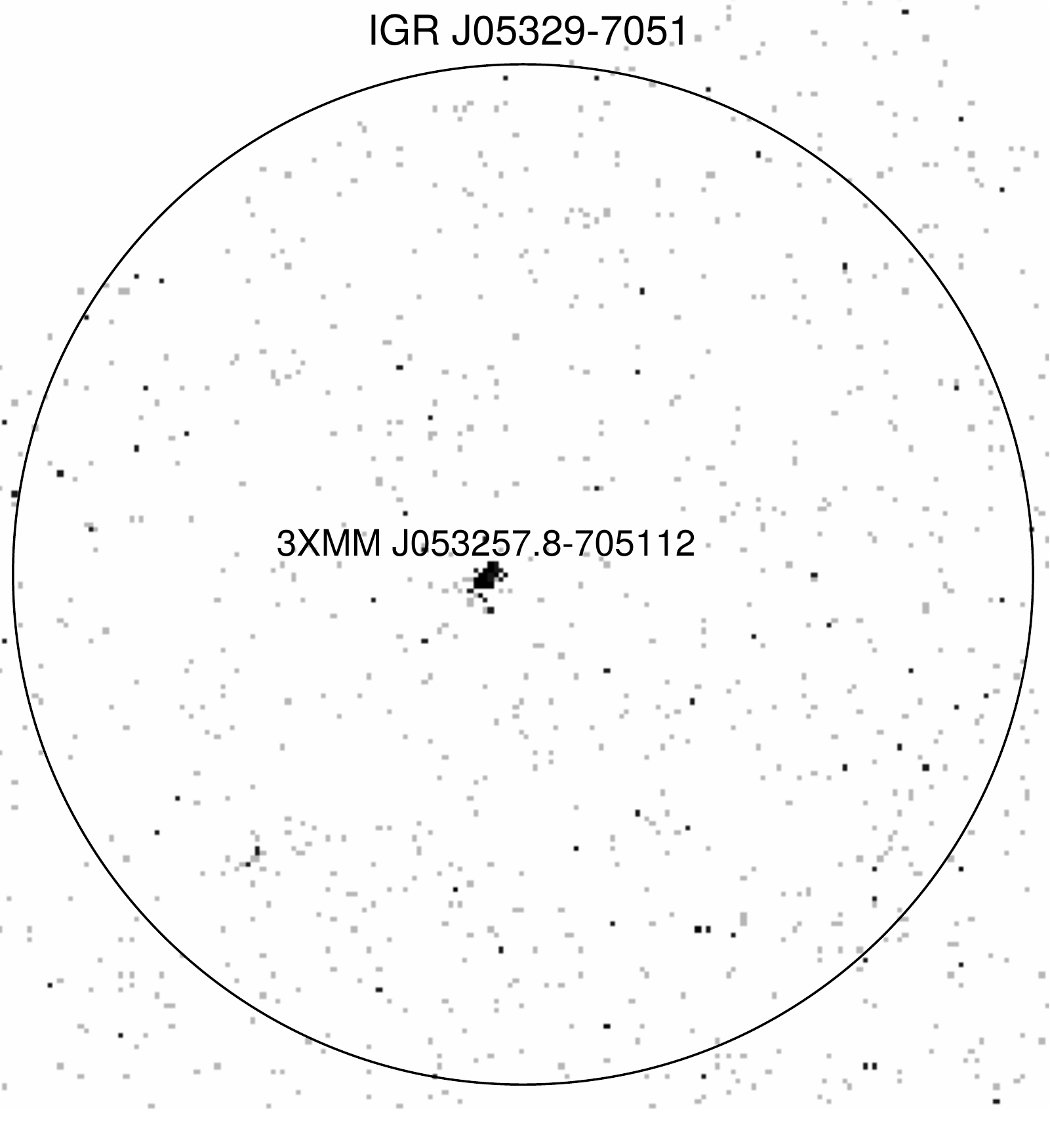}
\caption{{\it XMM- Newton} EPIC MOS 0.2--10~keV image of the field around IGR\,J05329$-$7051. The circle denotes the {\it INTEGRAL} error region of 4.2\arcmin\ in radius.}
\label{fig:igrj05329}
\end{figure}

The error circle of IGR\,J05329$-$7051 contains one obvious soft X-ray counterpart -- 3XMM\,J053257.8$-$705112, located 20\arcsec\ away from the {\it INTEGRAL} position (Fig.~\ref{fig:igrj05329}) and having a flux of $\simeq2\times10^{-13}$~erg\,s$^{-1}$\,cm$^{-2}$ in the 0.2--12 keV band  \citep{rosen_15}. The optical counterpart of 3XMM\,J053257.8$-$705112 is a distant \citep[$z=1.238$,][]{kozlowski12} AGN, MQS\,J053258.11$-$705112.9. We extracted the source spectrum from the data of an {\it XMM-Newton} observation in October 2001 (ObsId 0089210901, exposure 22 ks). Fitting it with the {\sc phabs*zpowerlw} model from the {\sc XSPEC} package we obtained a low absorption column density $N_{H}=(1.7\pm0.8)\times 10^{21}$~cm$^{-2}$ consistent with the absorption in the Milky Way in this direction, and a moderate photon index of $2.2\pm0.4$. The corresponding model flux in the 0.2--10~keV energy range 
is $1.4\times10^{-13}$~erg~s$^{-1}$~cm$^{-2}$. Due to its high hard X-ray luminosity $L_{\rm 17-60~keV}\sim 3\times 10^{46}$~erg~s$^{-1}$ we classify this source as a candidate blazar.

\section*{IGR\,J13100+0830}

\begin{figure}
\centering
 \includegraphics[width=0.85\linewidth,bb=0 0 564 571]{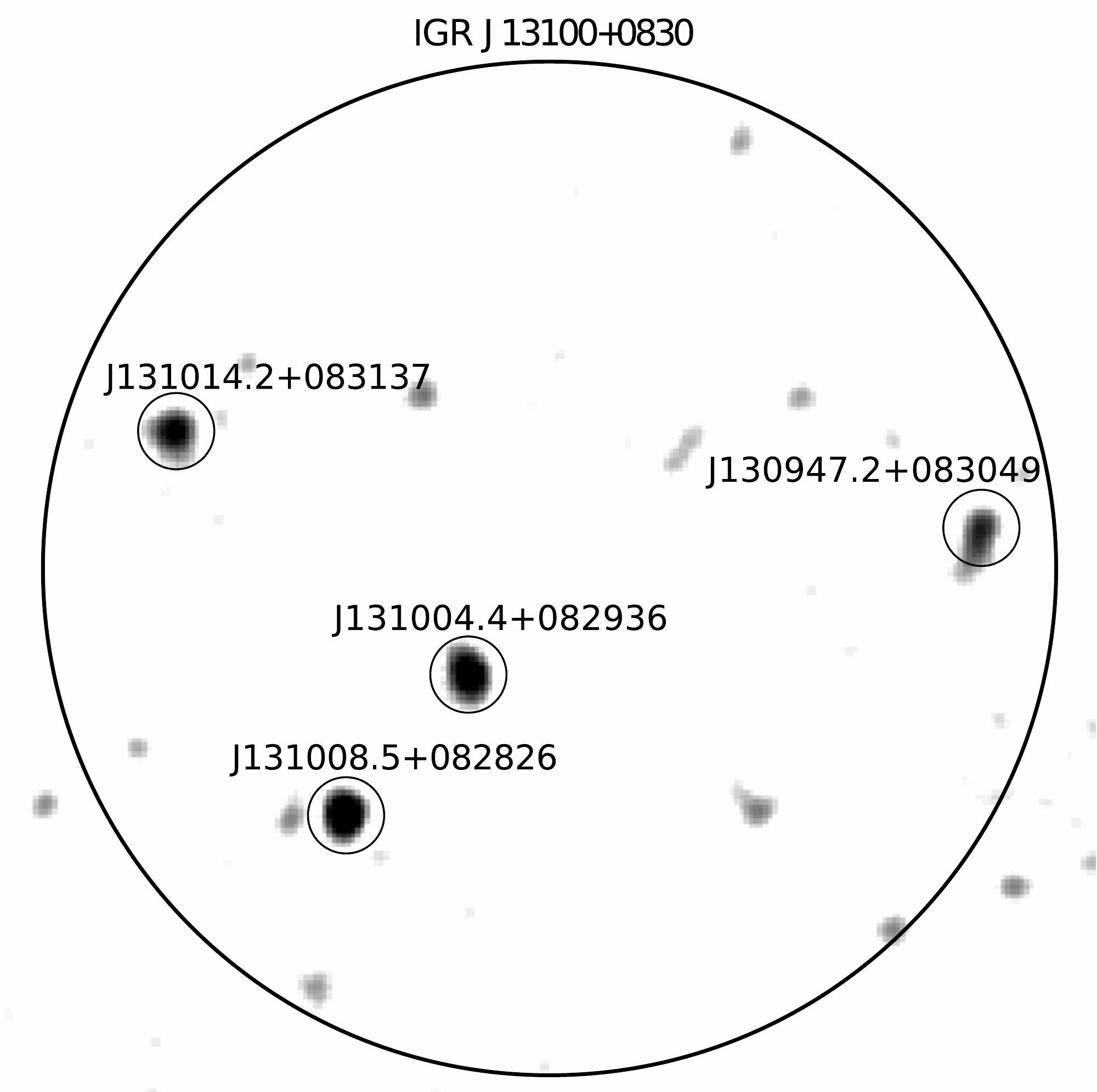}
 \caption{{\it Swift}/XRT 0.3--10~keV image of the field around IGR\,J13100+0830. The large circle of 4.2\arcmin\ in radius denotes the {\it INTEGRAL} error region. The smaller circles denote sources from the 1SXPS catalog \citep{evans_14}.} 
\label{fig:igrj13100}
\end{figure}

We found four soft X-ray counterparts in the 1SXPS catalog \citep{evans_14} within the $4.2'$ error cicrle of IGR\,J13100+0830, as shown in Fig.~\ref{fig:igrj13100}. Table~\ref{tab:13100srcs} lists offsets, count rates and optical counterparts for these objects.

Because of the highest 2--10~keV flux and nearest position to the {\it INTEGRAL} coordinates, we propose 1SXPS\,J131004.4+082936 as a probable counterpart, although contribution from other sources cannot be excluded. Observations with the {\it NuSTAR} hard X-ray focusing telescope \citep{harrison13_nustar} or the planned {\it Astro-H} mission \citep{takahashi10_ah} could help establish the nature of IGR\,J13100+0830 and find its optical counterpart.

\subsection{Catalog}
\label{sect:catalog}
The list of the detected sources with $S/N>4$ is presented in Table~\ref{tab:srcstab_red}, which consists of three blocks corresponding to the M81, LMC and 3C 273/Coma fields. The columns of the table are described below.

{\it Column (1) ''Id''} -- source number in the catalog.

{\it Column (2) ''Name''} -- source name. For sources previously detected in hard X-rays we use their catalog or common name. We assign an ``IGR'' name for sources detected for the first time (also highlighed in bold).

{\it Columns (3,4) ''R.A., Dec.''} -- right ascension and declination in equatorial coordinates (J2000 epoch).

{\it Column (5) ''S/N''} -- signal-to-noise ratio of the detected source.

{\it Column (6) ''Flux''} -- average source flux (17--60 keV) in mCrab and the associated 1$\sigma$ error.

{\it Columns (7,8) $D$, $z$} -- metric distance or redshift for extragalactic sources. For the calculation of luminosities (column 9) we used the metric distance for nearby sources ($z\le0.01$) and the luminosity distance estimated from the redshift for the more distant sources. Distances and redshifts were obtained from the SIMBAD and NED databases.

{\it Column (9) $\log L$} -- the logarithm of the 17--60~keV luminosity of the source. We only calculated luminosities for sources classified as AGN; a standard $\Lambda$CDM cosmology with $H_{0} = 67.8$ km s$^{-1}$ Mpc$^{-1}$, $\Omega_{m} = 0.308$ was used.

{\it Column (10) ''Type''} -- astrophysical type of the object: HMXB (LMXB) -- high(low)-mass X-ray binary; CV -- cataclysmic variable; pulsar -- rotation powered X-ray pulsar; cluster -- cluster of galaxies; Sy1, Sy2 (and intermediate types Sy1.2, Sy1.5, Sy1.8, Sy1.9) -- Seyfert galaxies of different types; NLS1 -- narrow-line Seyfert 1 galaxies; LINER -- low ionization nuclear emission-line region galaxy; XBONG -- X-ray bright optically normal galaxy; blazar -- BL Lac object or flat-spectrum radio quasar; NLRG - narrow emission-line radio-galaxy. For all sources associated with galaxies but without known activity type we ascribe an "AGN" type.

We should note, that there are few sources which classificated as LINER based on optical observations but shows unusually high hard X-ray luminosities - more than 10$^{43}$ erg\,s$^{-1}$, we decided to denote them as ``LINER?'.'  

{\it Column (11) ''Notes''} -- For known sources, we present an optical or IR counterpart name. For sources detected for the first time, we specify the soft X-ray counterpart and associated optical association. Some additional remarks are also provided.

\begin{figure*}
 \includegraphics[width=0.75\linewidth,bb=0 0 860 570]{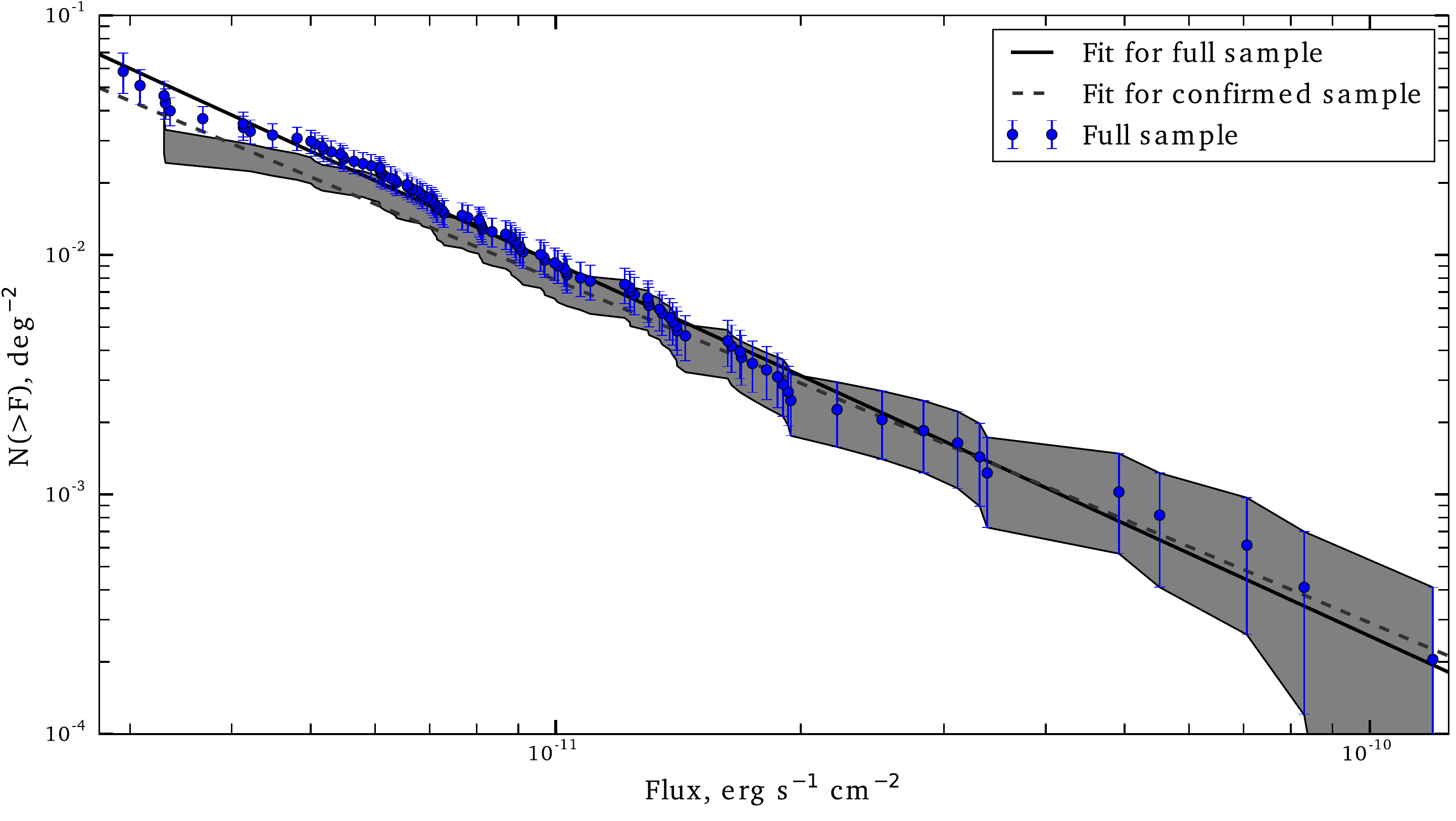}
 \caption{
   Number--flux (17--60~keV) relation for AGN. Blue points 
   represent the full AGN sample (80 confirmed non-blazar AGN and 25
   unidentified sources), while the black solid line shows the
   corresponding best-fitting power law model (the best-fit parameters
   are given in Table~\ref{tab:logn}). The shaded area represents the
   1$\sigma$ error region for the confirmed AGN sample composed of 80
   non-blazar AGN. The power-law fit for this sample is shown by the
   gray dashed line.} 
\label{fig:lnlstot}
\end{figure*}
\begin{figure*}
 \includegraphics[width=0.98\linewidth]{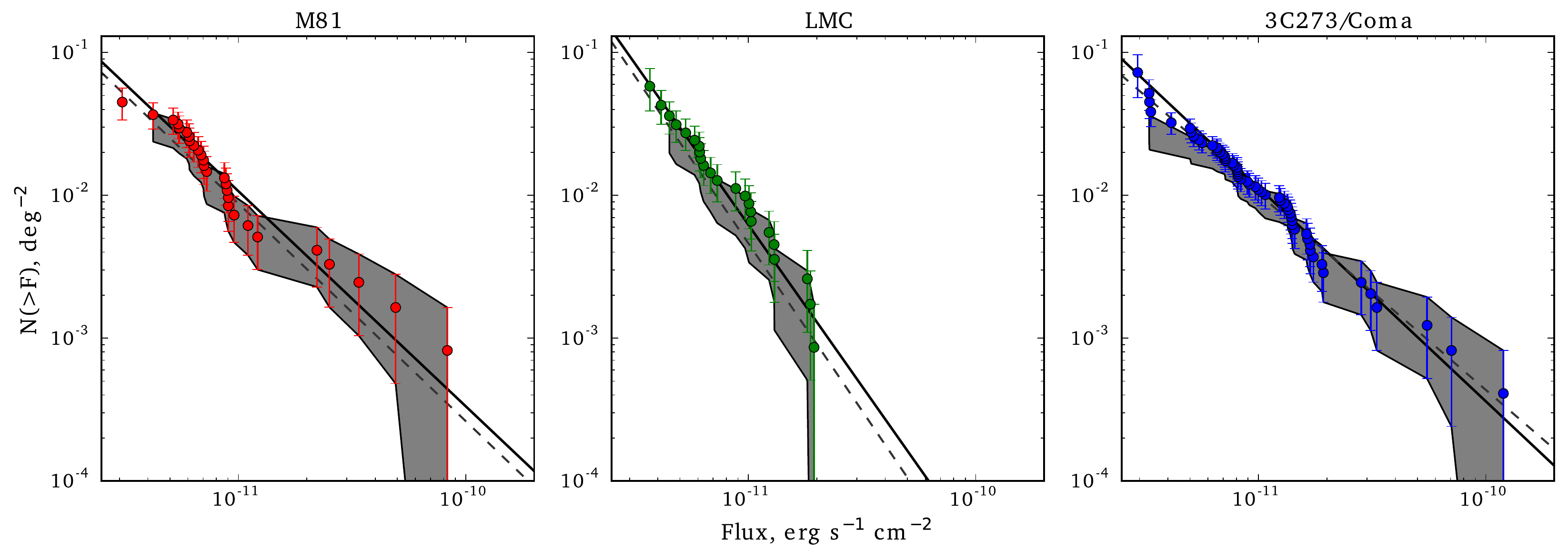}
 \caption{
   Number--flux (17--60~keV) relations for sources in the three 
   extragalactic fields. The colored dots show both known non-blazar
   AGN and unidentified sources, and the black solid lines represent
   the corresponding power-law fits (the parameter values are given in
   Table~\ref{tab:logn}). The shaded areas represent the 1$\sigma$
   regions of best-fit parameters excluding unidentified
   sources. Power-law fits for samples without unidentified sources
   are shown as dashed gray lines.}
\label{fig:lognlogs}
\end{figure*}			

\section{AGN sample and statistics}

Our resulting source catalog is dominated by extragalactic objects: the total sample of 98 AGN includes 64 Seyfert galaxies, 7 LINERs, 3 XBONGs, 16 blazars (or candidate blazars) and 8 AGN of unclear type. The catalog also contains 25 unidentified sources, thus the survey's identification is complete at 83\%. 

The {\it INTEGRAL}/IBIS deep extragalactic survey can be used to construct a number-flux relation for AGN, assuming that they are uniformly distributed in space (we check this assumption below). To this end, we excluded from the full AGN sample M81, NGC\,4151 and 3C\,273, since these were dedicated targets of long {\it INTEGRAL} observations, and 15 (3C\,273 already excluded) blazars. The resulting sample of 80 confirmed non-blazar AGN is referred to as a {\it confirmed AGN sample} hereafter. 

Fig.\,\ref{fig:lnlstot} shows the cumulative $\log{N}$--$\log{S}$ distribution derived from the confirmed AGN sample and corrected for the survey's sky coverage (Fig.~\ref{fig:coverage}). This distribution can be well described by a power law $N(>S) = AS^{-\alpha}$. Using the maximum likelihood estimator \citep{crawford70_ml} and the source number counts, we determined the slope $\alpha=1.44\pm0.14$ and normalization $A=(2.9\pm0.3)\times10^{-3}$~deg$^{-2}$, at the flux $2\times10^{-11}$~erg~s$^{-1}$~cm$^{-2}$, of the number-flux relation. Although the derived slope is consistent with that (3/2) expected for homogeneously distributed objects, there is an indication of some flattening of the $\log{N}$--$\log{S}$ distribution below $\simeq$6$\times$10$^{-12}$ erg s$^{-1}$ cm$^{-2}$, which may be caused by the incompleteness of our survey at these low fluxes.

Taking into account the high Galactic latitudes of the extragalactic fields under consideration, it is reasonable to expect that most of the unidentified sources in our sample have an extragalactic 
nature. We therefore also constructed a larger {\it full AGN sample} by adding all 25 unidentified sources to our confirmed AGN sample. The new sample includes 105 hard X-ray sources 
spanning down to a flux of $\simeq$3$\times$10$^{-12}$ erg\,s$^{-1}$\,cm$^{-2}$, which is a factor of two deeper than the all-sky extragalactic $\log{N}$--$\log{S}$ relation 
constructed by \citet{krivonos10_as2}. Using the same approach as before, we derived the best-fit slope $\alpha=1.56\pm0.13$ and normalization 
$A=(3.1\pm0.3)\times10^{-3}$~deg$^{-2}$ at the flux of $2\times10^{-11}$~erg~s$^{-1}$~cm$^{-2}$ for the number-flux relation constructed from the 
full AGN sample (see Fig.~\ref{fig:lnlstot} and Table~\ref{tab:logn}). These values are not significantly different from our estimates based on the confirmed AGN sample. The derived slope 
is consistent with 3/2, while the normalization is slightly lower but still consistent with the value of $(3.59\pm0.35)\times10^{-3}$~deg$^{-2}$ obtained 
for the all-sky {\it INTEGRAL} survey by \citet{krivonos10_as2}.

We further used the full AGN sample to construct the $\log{N}$--$\log{S}$ distributions for the three individual extragalactic fields (Fig.\,\ref{fig:lognlogs}). The corresponding 
slopes and normalizations are summarized in Table~\ref{tab:logn}. The slopes and normalizations for the M81 and 3C~273/Coma fields are compatible with each other while 
the LMC field shows a significantly steeper slope, yet consistent within 2$\sigma$ with the $\log{N}$--$\log{S}$ fit for the combined survey. The apparent lack 
of bright ($>2\times$10$^{-11}$ erg s$^{-1}$ cm$^{-2}$) AGN in the direction of the Large Magellanic Cloud was previously noticed by \citet{lutovinov12_agn}.

\begin{table*}
\caption{Best--fit parameters for the number-flux relations and estimated AGN space densities.}
\label{tab:logn}
\begin{minipage}[th]{\textwidth}
\centering
\begin{tabular}{|c|c|c|c|c|c|} \hline
Parameter & Units &  M81 & LMC & 3C~273/Coma & Total \\
\hline
\multicolumn{6}{c}{$\log{N}$--$\log{S}$, best-fit parameters for full
  AGN sample (confirmed AGN and unidentified sources)}\\
\hline
$\alpha$ &  & 1.51$\pm$0.23 & 2.26$\pm$0.36  & 1.50$\pm$0.17  & $1.56\pm0.13$ \\
$A^{1}$ & $\times10^{-3}$~deg$^{-2}$ & 3.8$\pm$0.7 & 1.3$\pm$0.3 & 4.0$\pm$0.5 &  $3.1\pm0.3$ \\
$N^{2}$ & & 27 & 23 & 55 &  105 \\
\hline
\multicolumn{6}{c}{$\log{N}$--$\log{S}$, best-fit parameters for confirmed AGN sample }\\
\hline
$\alpha$ &  & 1.52$\pm$0.27 & 2.34$\pm$0.46  & 1.37$\pm$0.19  & $1.43\pm0.14$ \\
$A^{1}$ & $\times10^{-3}$~deg$^{-2}$ & 3.0$\pm$0.6 & 0.9$\pm$0.2 & 4.0$\pm$0.6 &  $2.9\pm0.3$ \\
$N^{3}$ & & 22 & 16 & 42 &  80 \\
\hline
$N^{4}$ & & 12 & 5 & 27 &  44 \\
$N^{5}$ & & 9 & 4 & 18 &  31 \\
\hline
\multicolumn{6}{c}{AGN space density estimated by the 1/V$_{max}$ method} \\
\hline
$\rho^{6}$ & $\times10^{-5}$~Mpc$^{-3}$ & 14.8$\pm$11.6& 3.2$\pm$2.1& 5.4 $\pm$ 2.1& 7.6 $\pm$ 3.5\\
\hline
\end{tabular}
\end{minipage}
\begin{flushleft}
$^{1}$ The normalization $A$ is derived at the flux $2\times10^{-11}$~erg~s$^{-1}$~cm$^{-2}$.\\
$^{2}$ The number of confirmed non-blazar AGN and unidentified sources
  in the field.\\
$^{3}$ The number of confirmed non-blazar AGN in the field.\\
$^{4}$ The number of confirmed non-blazar AGN at $D<150$~Mpc.\\
$^{5}$ The number of confirmed non-blazar AGN at $D<150$~Mpc with $L>10^{42}$~erg s$^{-1}$.\\
$^{6}$ Number density of AGN at $D<150$~Mpc with $L>10^{42}$~erg s$^{-1}$.
\end{flushleft}
\end{table*}

\begin{figure}
 \includegraphics[width=\linewidth,bb=0 0 502 502,clip]{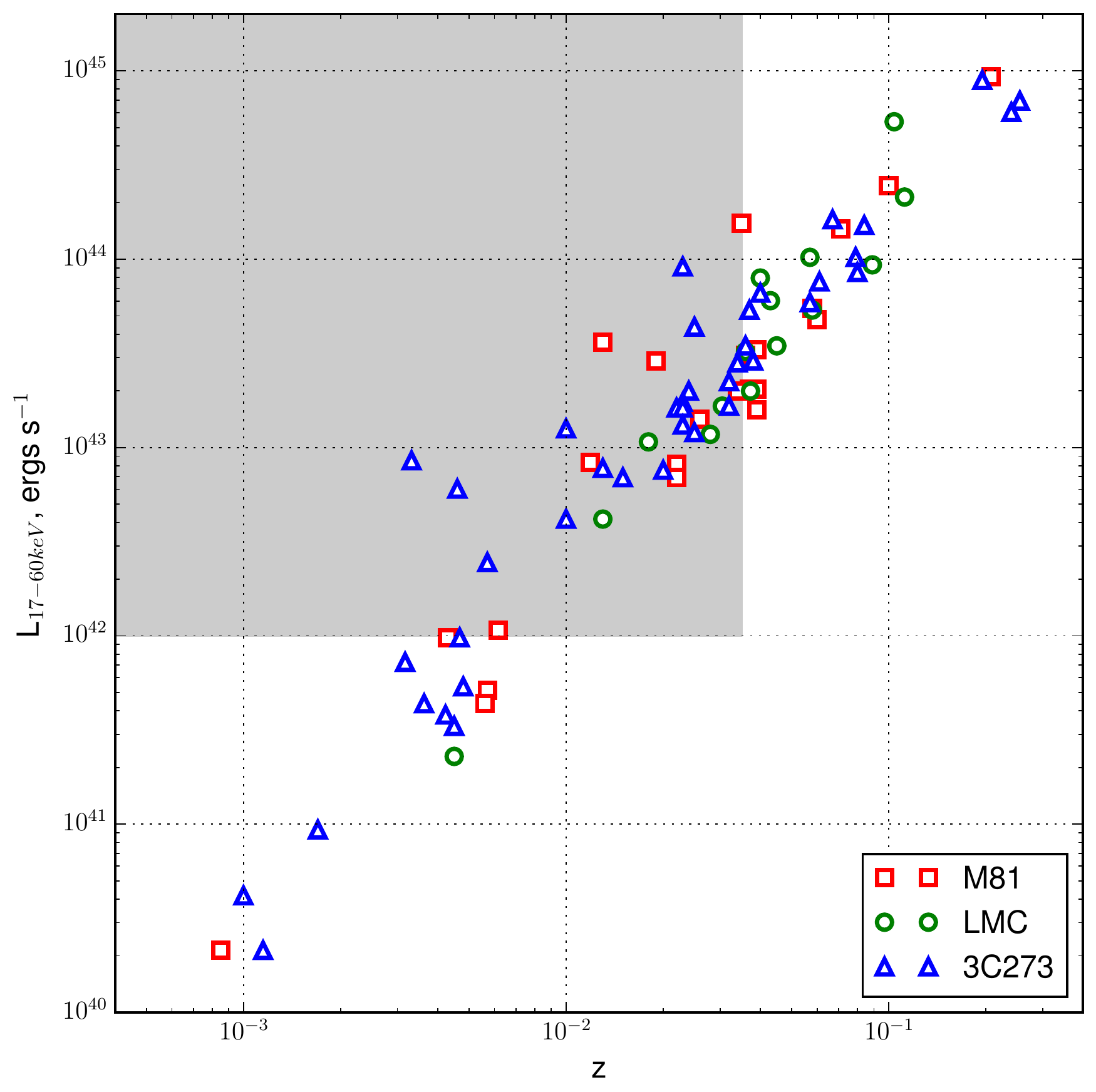}
 \caption{Hard X-ray luminosty vs. redshift for the identified non-blazar AGN. The sources from shaded region were used for estimating the AGN space density (see text for details). }
 \label{fig:z_L}
\end{figure}

It is well known that the spatial distribution of galaxies in the local Universe is strongly inhomogeneous on scales less than $\sim100$--200~Mpc \citep{jarrett04}. With the typical sensitivity of our deep {\it INTEGRAL}/IBIS survey $\sim 5\times 10^{-12}$~erg~s$^{-1}$~cm$^{-2}$ (17--60~keV) we can detect AGN with the characteristic luminosity of $L_{*}\sim 5\times 10^{43}$~erg~s$^{-1}$ \citep{sazonov07_agn, sazonov15_obsagn} beyond this scale, out to $\sim 300$~Mpc. This fact is confirmed by Fig.\,\ref{fig:z_L}, where we plot the hard X-ray luminosity vs. redshift for the confirmed AGN sample. Therefore, we can make use of the AGN count statistics in the directions to M81, LMC and 3C~273/Coma fields to probe the matter distribution in the local Universe regarding AGN as its tracers  \citep{krivonos07_allsky,ajello12_bat60}. 

To get a rough idea, we can compare the numbers of nearby AGN (at distances $D<150$~Mpc) found in the three fields (see Table~\ref{tab:logn}). There are 12 such objects (i.e. confirmed non-blazar AGN) in the M\,81 field, 5 in the LMC field and 27 in the 3C\,273/Coma field. Given that the last field is two times larger than the former two, these numbers do not indicate a significant difference in the AGN space density in the three considered directions. 
More accurate estimates can be achieved with the {\it 1/V$_{max}$} method \citep{schmidt68,huchra73}. We restricted our analysis to sources with 
$D<150$~Mpc and L$>$10$^{42}$ erg s$^{-1}$ (within the shaded region in Fig.~\ref{fig:z_L}). The latter condition is imposed to diminish the statistical noise associated with the lowest luminosity (and hence very nearby) AGN. The resulting samples contain 9, 4 and 18 AGN in the M\,81, LMC and 3C\,273/Coma fields, respectively. As can be seen from Table~\ref{tab:logn}, the LMC field exhibits the lowest local AGN space density. However, the estimated densities for the different fields are consistent with the density estimated by combining these fields.

\section{Summary}

We have analyzed the deepest {\it INTEGRAL} hard X-ray survey of three extragalactic fields: around M81, LMC and 3C\,273/Coma, with the peak exposure of 6.8 Ms in the LMC field and above 9 Ms in the two other fields. The peak achieved sensitivity is 0.18~mCrab in the 17--60~keV energy band. The catalog of sources detected in the combined survey contains 147 objects detected above the threshold of $S/N>4$, with 37 of them having been detected in hard X-rays for the first time. We have identified 13 of the newly detected objects using archival soft X-ray observations. Twenty five sources (24 of new and SWIFT J0826.2--7033) remain unidentified making the completeness of the survey at the level of 83\%.

The catalog is dominated by extragalactic sources. The cumulative $\log{N}$--$\log{S}$ distribution of non-blazar AGN is consistent with a power law down 
to fluxes $\simeq3\times10^{-12}$ erg\,s$^{-1}$\,cm$^{-2}$, which is deeper by a factor of two compared to the previous (all-sky) measurement of \citet{krivonos10_as2}. The 
AGN number counts for the M81 and 3C 273/Coma fields are consistent with each other, while the LMC field demonstrates a steeper number-flux distribution 
(2$\sigma$ deviation from the expected $-3/2$ slope)  and a lack of bright AGN with flux higher than 2$\times$10$^{-11}$ erg s$^{-1}$ cm$^{-2}$.

\section*{Acknowledgements}
This work is based on observations with {\it INTEGRAL}, an ESA project with instruments and the science data centre funded by ESA member states (especially the PI countries: Denmark, France, Germany, Italy, Switzerland, Spain), and Poland, and with the participation of Russia and the USA. The data were obtained from the European\footnote{\url{http://isdc.unige.ch}} and Russian\footnote{\url{http://hea.iki.rssi.ru/rsdc}} {\it INTEGRAL} Science Data Centers. This research was supported by the Russian Science Foundation (grant 14-22-00271). The authors thank the Max-Planck-Institut f\"{u}r Astrophysik for computational support.
\bibliographystyle{mnras}
\bibliography{igrbib}

\begin{thebibliography}{}
\makeatletter
\relax
\def\mn@urlcharsother{\let\do\@makeother \do\$\do\&\do\#\do\^\do\_\do\%\do\~}
\def\mn@doi{\begingroup\mn@urlcharsother \@ifnextchar [ {\mn@doi@}
  {\mn@doi@[]}}
\def\mn@doi@[#1]#2{\def\@tempa{#1}\ifx\@tempa\@empty \href
  {http://dx.doi.org/#2} {doi:#2}\else \href {http://dx.doi.org/#2} {#1}\fi
  \endgroup}
\def\mn@eprint#1#2{\mn@eprint@#1:#2::\@nil}
\def\mn@eprint@arXiv#1{\href {http://arxiv.org/abs/#1} {{\tt arXiv:#1}}}
\def\mn@eprint@dblp#1{\href {http://dblp.uni-trier.de/rec/bibtex/#1.xml}
  {dblp:#1}}
\def\mn@eprint@#1:#2:#3:#4\@nil{\def\@tempa {#1}\def\@tempb {#2}\def\@tempc
  {#3}\ifx \@tempc \@empty \let \@tempc \@tempb \let \@tempb \@tempa \fi \ifx
  \@tempb \@empty \def\@tempb {arXiv}\fi \@ifundefined
  {mn@eprint@\@tempb}{\@tempb:\@tempc}{\expandafter \expandafter \csname
  mn@eprint@\@tempb\endcsname \expandafter{\@tempc}}}

\bibitem[\protect\citeauthoryear{{Ajello} et~al.,}{{Ajello}
  et~al.}{2009}]{ajello09_cl}
{Ajello} M.,  et~al., 2009, \mn@doi [ApJ] {10.1088/0004-637X/690/1/367}, \href
  {http://adsabs.harvard.edu/abs/2009ApJ...690..367A} {690, 367}

\bibitem[\protect\citeauthoryear{{Ajello}, {Alexander}, {Greiner}, {Madejski},
  {Gehrels}  \& {Burlon}}{{Ajello} et~al.}{2012}]{ajello12_bat60}
{Ajello} M.,  {Alexander} D.~M.,  {Greiner} J.,  {Madejski} G.~M.,  {Gehrels}
  N.,   {Burlon} D.,  2012, \mn@doi [ApJ] {10.1088/0004-637X/749/1/21}, \href
  {http://adsabs.harvard.edu/abs/2012ApJ...749...21A} {749, 21}

\bibitem[\protect\citeauthoryear{{Barlow}, {Knigge}, {Bird}, {J Dean}, {Clark},
  {Hill}, {Molina}  \& {Sguera}}{{Barlow} et~al.}{2006}]{barlow06}
{Barlow} E.~J.,  {Knigge} C.,  {Bird} A.~J.,  {J Dean} A.,  {Clark} D.~J.,
  {Hill} A.~B.,  {Molina} M.,   {Sguera} V.,  2006, \mn@doi [MNRAS]
  {10.1111/j.1365-2966.2006.10836.x}, \href
  {http://adsabs.harvard.edu/abs/2006MNRAS.372..224B} {372, 224}

\bibitem[\protect\citeauthoryear{{Baumgartner}, {Tueller}, {Markwardt},
  {Skinner}, {Barthelmy}, {Mushotzky}, {Evans}  \& {Gehrels}}{{Baumgartner}
  et~al.}{2013}]{baumgartner13_swift70}
{Baumgartner} W.~H.,  {Tueller} J.,  {Markwardt} C.~B.,  {Skinner} G.~K.,
  {Barthelmy} S.,  {Mushotzky} R.~F.,  {Evans} P.~A.,   {Gehrels} N.,  2013,
  \mn@doi [ApJS] {10.1088/0067-0049/207/2/19}, \href
  {http://adsabs.harvard.edu/abs/2013ApJS..207...19B} {207, 19}

\bibitem[\protect\citeauthoryear{{Beckmann} et~al.,}{{Beckmann}
  et~al.}{2009}]{beckmann09}
{Beckmann} V.,  et~al., 2009, \mn@doi [A\&A] {10.1051/0004-6361/200912111},
  \href {http://adsabs.harvard.edu/abs/2009A%26A...505..417B} {505, 417}

\bibitem[\protect\citeauthoryear{{Bird} et~al.,}{{Bird} et~al.}{2010}]{bird09}
{Bird} A.~J.,  et~al., 2010, \mn@doi [ApJ] {10.1088/0067-0049/186/1/1}, \href
  {http://adsabs.harvard.edu/abs/2010ApJS..186....1B} {186, 1}

\bibitem[\protect\citeauthoryear{{Bird} et~al.,}{{Bird} et~al.}{2016}]{bird16}
{Bird} A.~J.,  et~al., 2016, preprint, \href
  {http://adsabs.harvard.edu/abs/2016arXiv160106074B} {} (\mn@eprint {arXiv}
  {1601.06074})

\bibitem[\protect\citeauthoryear{{Bodaghee}, {Tomsick}, {Rodriguez}  \&
  {James}}{{Bodaghee} et~al.}{2012}]{bodaghee12}
{Bodaghee} A.,  {Tomsick} J.~A.,  {Rodriguez} J.,   {James} J.~B.,  2012,
  \mn@doi [ApJ] {10.1088/0004-637X/744/2/108}, \href
  {http://adsabs.harvard.edu/abs/2012ApJ...744..108B} {744, 108}

\bibitem[\protect\citeauthoryear{{Brandt} \& {Alexander}}{{Brandt} \&
  {Alexander}}{2015}]{brandt15_rev}
{Brandt} W.~N.,  {Alexander} D.~M.,  2015, \mn@doi [AApR]
  {10.1007/s00159-014-0081-z}, \href
  {http://adsabs.harvard.edu/abs/2015A%26ARv..23....1B} {23, 1}

\bibitem[\protect\citeauthoryear{{Caballero} et~al.,}{{Caballero}
  et~al.}{2013}]{caballero13}
{Caballero} I.,  et~al., 2013, preprint, \href
  {http://adsabs.harvard.edu/abs/2013arXiv1304.1349C} {} (\mn@eprint {arXiv}
  {1304.1349})

\bibitem[\protect\citeauthoryear{{Churazov} et~al.,}{{Churazov}
  et~al.}{2014}]{churazov14}
{Churazov} E.,  et~al., 2014, \mn@doi [Nature] {10.1038/nature13672}, \href
  {http://adsabs.harvard.edu/abs/2014Natur.512..406C} {512, 406}

\bibitem[\protect\citeauthoryear{{Crawford}, {Jauncey}  \&
  {Murdoch}}{{Crawford} et~al.}{1970}]{crawford70_ml}
{Crawford} D.~F.,  {Jauncey} D.~L.,   {Murdoch} H.~S.,  1970, \mn@doi [ApJ]
  {10.1086/150672}, \href {http://adsabs.harvard.edu/abs/1970ApJ...162..405C}
  {162, 405}

\bibitem[\protect\citeauthoryear{{Cusumano} et~al.,}{{Cusumano}
  et~al.}{2010}]{cusumano10}
{Cusumano} G.,  et~al., 2010, \mn@doi [A\&A] {10.1051/0004-6361/201015249},
  \href {http://adsabs.harvard.edu/abs/2010A%26A...524A..64C} {524, A64}

\bibitem[\protect\citeauthoryear{{ESA}}{{ESA}}{1997}]{tycho_cat}
{ESA} ed. 1997, {The HIPPARCOS and TYCHO catalogues. Astrometric and
  photometric star catalogues derived from the ESA HIPPARCOS Space Astrometry
  Mission}  ESA Special Publication Vol. 1200

\bibitem[\protect\citeauthoryear{Evans et~al.,}{Evans et~al.}{2014}]{evans_14}
Evans P.~A.,  et~al., 2014, ApJS, 210, 8

\bibitem[\protect\citeauthoryear{{Gehrels} et~al.,}{{Gehrels}
  et~al.}{2004}]{gehrels04_swift}
{Gehrels} N.,  et~al., 2004, \mn@doi [ApJ] {10.1086/422091}, \href
  {http://adsabs.harvard.edu/abs/2004ApJ...611.1005G} {611, 1005}

\bibitem[\protect\citeauthoryear{{G{\"o}tz}, {Mereghetti}, {Merlini}, {Sidoli}
  \& {Belloni}}{{G{\"o}tz} et~al.}{2006}]{gotz06}
{G{\"o}tz} D.,  {Mereghetti} S.,  {Merlini} D.,  {Sidoli} L.,   {Belloni} T.,
  2006, \mn@doi [A\&A] {10.1051/0004-6361:20053744}, \href
  {http://adsabs.harvard.edu/abs/2006A%26A...448..873G} {448, 873}

\bibitem[\protect\citeauthoryear{{Grebenev}, {Lutovinov}, {Tsygankov}  \&
  {Winkler}}{{Grebenev} et~al.}{2012}]{grebenev_12sn}
{Grebenev} S.~A.,  {Lutovinov} A.~A.,  {Tsygankov} S.~S.,   {Winkler} C.,
  2012, \mn@doi [Nature] {10.1038/nature11473}, \href
  {http://adsabs.harvard.edu/abs/2012Natur.490..373G} {490, 373}

\bibitem[\protect\citeauthoryear{{Grebenev}, {Lutovinov}, {Tsygankov}  \&
  {Mereminskiy}}{{Grebenev} et~al.}{2013}]{grebenev13LMC}
{Grebenev} S.~A.,  {Lutovinov} A.~A.,  {Tsygankov} S.~S.,   {Mereminskiy}
  I.~A.,  2013, \mn@doi [MNRAS] {10.1093/mnras/sts008}, \href
  {http://adsabs.harvard.edu/abs/2013MNRAS.428...50G} {428, 50}

\bibitem[\protect\citeauthoryear{{Harrison} et~al.,}{{Harrison}
  et~al.}{2013}]{harrison13_nustar}
{Harrison} F.~A.,  et~al., 2013, \mn@doi [ApJ] {10.1088/0004-637X/770/2/103},
  \href {http://adsabs.harvard.edu/abs/2013ApJ...770..103H} {770, 103}

\bibitem[\protect\citeauthoryear{{Huchra} \& {Sargent}}{{Huchra} \&
  {Sargent}}{1973}]{huchra73}
{Huchra} J.,  {Sargent} W.~L.~W.,  1973, \mn@doi [ApJ] {10.1086/152510}, \href
  {http://adsabs.harvard.edu/abs/1973ApJ...186..433H} {186, 433}

\bibitem[\protect\citeauthoryear{{Jarrett}}{{Jarrett}}{2004}]{jarrett04}
{Jarrett} T.,  2004, \mn@doi [PASA] {10.1071/AS04050}, \href
  {http://adsabs.harvard.edu/abs/2004PASA...21..396J} {21, 396}

\bibitem[\protect\citeauthoryear{{Koz{\l}owski} et~al.,}{{Koz{\l}owski}
  et~al.}{2012}]{kozlowski12}
{Koz{\l}owski} S.,  et~al., 2012, \mn@doi [ApJ] {10.1088/0004-637X/746/1/27},
  \href {http://cdsads.u-strasbg.fr/abs/2012ApJ...746...27K} {746, 27}

\bibitem[\protect\citeauthoryear{{Krivonos}, {Vikhlinin}, {Churazov},
  {Lutovinov}, {Molkov}  \& {Sunyaev}}{{Krivonos}
  et~al.}{2005}]{krivonos05_coma}
{Krivonos} R.,  {Vikhlinin} A.,  {Churazov} E.,  {Lutovinov} A.,  {Molkov} S.,
   {Sunyaev} R.,  2005, \mn@doi [ApJ] {10.1086/429657}, \href
  {http://adsabs.harvard.edu/abs/2005ApJ...625...89K} {625, 89}

\bibitem[\protect\citeauthoryear{{Krivonos}, {Revnivtsev}, {Lutovinov},
  {Sazonov}, {Churazov}  \& {Sunyaev}}{{Krivonos}
  et~al.}{2007}]{krivonos07_allsky}
{Krivonos} R.,  {Revnivtsev} M.,  {Lutovinov} A.,  {Sazonov} S.,  {Churazov}
  E.,   {Sunyaev} R.,  2007, \mn@doi [A\&A] {10.1051/0004-6361:20077191}, \href
  {http://adsabs.harvard.edu/abs/2007A%26A...475..775K} {475, 775}

\bibitem[\protect\citeauthoryear{{Krivonos}, {Revnivtsev}, {Tsygankov},
  {Sazonov}, {Vikhlinin}, {Pavlinsky}, {Churazov}  \& {Sunyaev}}{{Krivonos}
  et~al.}{2010a}]{krivonos10_as1}
{Krivonos} R.,  {Revnivtsev} M.,  {Tsygankov} S.,  {Sazonov} S.,  {Vikhlinin}
  A.,  {Pavlinsky} M.,  {Churazov} E.,   {Sunyaev} R.,  2010a, \mn@doi [A\&A]
  {10.1051/0004-6361/200913814}, \href
  {http://adsabs.harvard.edu/abs/2010A%26A...519A.107K} {519, A107}

\bibitem[\protect\citeauthoryear{{Krivonos}, {Tsygankov}, {Revnivtsev},
  {Grebenev}, {Churazov}  \& {Sunyaev}}{{Krivonos}
  et~al.}{2010b}]{krivonos10_as2}
{Krivonos} R.,  {Tsygankov} S.,  {Revnivtsev} M.,  {Grebenev} S.,  {Churazov}
  E.,   {Sunyaev} R.,  2010b, \mn@doi [A\&A] {10.1051/0004-6361/201014935},
  \href {http://adsabs.harvard.edu/abs/2010A%26A...523A..61K} {523, A61}

\bibitem[\protect\citeauthoryear{{Krivonos}, {Tsygankov}, {Lutovinov},
  {Revnivtsev}, {Churazov}  \& {Sunyaev}}{{Krivonos} et~al.}{2012}]{krivonos12}
{Krivonos} R.,  {Tsygankov} S.,  {Lutovinov} A.,  {Revnivtsev} M.,  {Churazov}
  E.,   {Sunyaev} R.,  2012, \mn@doi [A\&A] {10.1051/0004-6361/201219617},
  \href {http://adsabs.harvard.edu/abs/2012A%26A...545A..27K} {545, A27}

\bibitem[\protect\citeauthoryear{{Lutovinov}, {Vikhlinin}, {Churazov},
  {Revnivtsev}  \& {Sunyaev}}{{Lutovinov} et~al.}{2008}]{lutovinov08_coma}
{Lutovinov} A.~A.,  {Vikhlinin} A.,  {Churazov} E.~M.,  {Revnivtsev} M.~G.,
  {Sunyaev} R.~A.,  2008, \mn@doi [ApJ] {10.1086/592032}, \href
  {http://adsabs.harvard.edu/abs/2008ApJ...687..968L} {687, 968}

\bibitem[\protect\citeauthoryear{{Lutovinov}, {Grebenev}  \&
  {Tsygankov}}{{Lutovinov} et~al.}{2012}]{lutovinov12_agn}
{Lutovinov} A.~A.,  {Grebenev} S.~A.,   {Tsygankov} S.~S.,  2012, \mn@doi
  [Astronomy Letters] {10.1134/S1063773712080051}, \href
  {http://adsabs.harvard.edu/abs/2012AstL...38..492L} {38, 492}

\bibitem[\protect\citeauthoryear{{Lutovinov}, {Revnivtsev}, {Tsygankov}  \&
  {Krivonos}}{{Lutovinov} et~al.}{2013}]{lutovinov13}
{Lutovinov} A.~A.,  {Revnivtsev} M.~G.,  {Tsygankov} S.~S.,   {Krivonos} R.~A.,
   2013, \mn@doi [MNRAS] {10.1093/mnras/stt168}, \href
  {http://adsabs.harvard.edu/abs/2013MNRAS.431..327L} {431, 327}

\bibitem[\protect\citeauthoryear{{Malizia}, {Stephen}, {Bassani}, {Bird},
  {Panessa}  \& {Ubertini}}{{Malizia} et~al.}{2009}]{malizia09}
{Malizia} A.,  {Stephen} J.~B.,  {Bassani} L.,  {Bird} A.~J.,  {Panessa} F.,
  {Ubertini} P.,  2009, \mn@doi [MNRAS] {10.1111/j.1365-2966.2009.15330.x},
  \href {http://adsabs.harvard.edu/abs/2009MNRAS.399..944M} {399, 944}

\bibitem[\protect\citeauthoryear{{Molkov}, {Lutovinov}  \& {Falanga}}{{Molkov}
  et~al.}{2015}]{molkov15}
{Molkov} S.~V.,  {Lutovinov} A.~A.,   {Falanga} M.,  2015, \mn@doi [Astronomy
  Letters] {10.1134/S1063773715100047}, \href
  {http://adsabs.harvard.edu/abs/2015AstL...41..562M} {41, 562}

\bibitem[\protect\citeauthoryear{Mullaney et~al.,}{Mullaney
  et~al.}{2015}]{mullaney15}
Mullaney J.~R.,  et~al., 2015, ApJ, 808, 184

\bibitem[\protect\citeauthoryear{{Paltani}, {Walter}, {McHardy}, {Dwelly},
  {Steiner}  \& {Courvoisier}}{{Paltani} et~al.}{2008}]{paltani08}
{Paltani} S.,  {Walter} R.,  {McHardy} I.~M.,  {Dwelly} T.,  {Steiner} C.,
  {Courvoisier} T.~J.-L.,  2008, \mn@doi [A\&A] {10.1051/0004-6361:200809450},
  \href {http://adsabs.harvard.edu/abs/2008A%26A...485..707P} {485, 707}

\bibitem[\protect\citeauthoryear{{Revnivtsev}, {Lutovinov}, {Churazov},
  {Sazonov}, {Gilfanov}, {Grebenev}  \& {Sunyaev}}{{Revnivtsev}
  et~al.}{2008}]{revnivtsev08}
{Revnivtsev} M.,  {Lutovinov} A.,  {Churazov} E.,  {Sazonov} S.,  {Gilfanov}
  M.,  {Grebenev} S.,   {Sunyaev} R.,  2008, \mn@doi [A\&A]
  {10.1051/0004-6361:200810115}, \href
  {http://adsabs.harvard.edu/abs/2008A%26A...491..209R} {491, 209}

\bibitem[\protect\citeauthoryear{{Rosen} et~al.,}{{Rosen}
  et~al.}{2015}]{rosen_15}
{Rosen} S.~R.,  et~al., 2015, preprint, \href
  {http://adsabs.harvard.edu/abs/2015arXiv150407051R} {} (\mn@eprint {arXiv}
  {1504.07051})

\bibitem[\protect\citeauthoryear{{Sazonov}, {Revnivtsev}, {Krivonos},
  {Churazov}  \& {Sunyaev}}{{Sazonov} et~al.}{2007}]{sazonov07_agn}
{Sazonov} S.,  {Revnivtsev} M.,  {Krivonos} R.,  {Churazov} E.,   {Sunyaev} R.,
   2007, \mn@doi [A\&A] {10.1051/0004-6361:20066277}, \href
  {http://adsabs.harvard.edu/abs/2007A%26A...462...57S} {462, 57}

\bibitem[\protect\citeauthoryear{{Sazonov}, {Krivonos}, {Revnivtsev},
  {Churazov}  \& {Sunyaev}}{{Sazonov} et~al.}{2008}]{sazonov08_agnspe}
{Sazonov} S.,  {Krivonos} R.,  {Revnivtsev} M.,  {Churazov} E.,   {Sunyaev} R.,
   2008, \mn@doi [A\&A] {10.1051/0004-6361:20078537}, \href
  {http://adsabs.harvard.edu/abs/2008A%26A...482..517S} {482, 517}

\bibitem[\protect\citeauthoryear{{Sazonov}, {Lutovinov}  \&
  {Krivonos}}{{Sazonov} et~al.}{2014}]{sazonov14_ulx}
{Sazonov} S.~Y.,  {Lutovinov} A.~A.,   {Krivonos} R.~A.,  2014, \mn@doi
  [Astronomy Letters] {10.1134/S1063773714030062}, \href
  {http://adsabs.harvard.edu/abs/2014AstL...40...65S} {40, 65}

\bibitem[\protect\citeauthoryear{{Sazonov}, {Churazov}  \&
  {Krivonos}}{{Sazonov} et~al.}{2015}]{sazonov15_obsagn}
{Sazonov} S.,  {Churazov} E.,   {Krivonos} R.,  2015, \mn@doi [MNRAS]
  {10.1093/mnras/stv2069}, \href
  {http://adsabs.harvard.edu/abs/2015MNRAS.454.1202S} {454, 1202}

\bibitem[\protect\citeauthoryear{{Schmidt}}{{Schmidt}}{1968}]{schmidt68}
{Schmidt} M.,  1968, \mn@doi [ApJ] {10.1086/149446}, \href
  {http://adsabs.harvard.edu/abs/1968ApJ...151..393S} {151, 393}

\bibitem[\protect\citeauthoryear{{Stickel} \& {Kuehr}}{{Stickel} \&
  {Kuehr}}{1993}]{stickel93}
{Stickel} M.,  {Kuehr} H.,  1993, A\&AS, \href
  {http://adsabs.harvard.edu/abs/1993A%26AS..100..395S} {100, 395}

\bibitem[\protect\citeauthoryear{{Takahashi} et~al.,}{{Takahashi}
  et~al.}{2010}]{takahashi10_ah}
{Takahashi} T.,  et~al., 2010, in Society of Photo-Optical Instrumentation
  Engineers (SPIE) Conference Series. p.~0 (\mn@eprint {arXiv} {1010.4972}),
  \mn@doi{10.1117/12.857875}

\bibitem[\protect\citeauthoryear{{Voges} et~al.,}{{Voges}
  et~al.}{1999}]{voges99}
{Voges} W.,  et~al., 1999, A\&A, \href
  {http://adsabs.harvard.edu/abs/1999A%26A...349..389V} {349, 389}

\bibitem[\protect\citeauthoryear{{Walter}, {Lutovinov}, {Bozzo}  \&
  {Tsygankov}}{{Walter} et~al.}{2015}]{walter15}
{Walter} R.,  {Lutovinov} A.~A.,  {Bozzo} E.,   {Tsygankov} S.~S.,  2015,
  \mn@doi [A\&ARv] {10.1007/s00159-015-0082-6}, \href
  {http://adsabs.harvard.edu/abs/2015A%26ARv..23....2W} {23, 2}

\bibitem[\protect\citeauthoryear{{Winkler} et~al.,}{{Winkler}
  et~al.}{2003}]{winkler03}
{Winkler} C.,  et~al., 2003, \mn@doi [A\&A] {10.1051/0004-6361:20031288}, \href
  {http://adsabs.harvard.edu/abs/2003A%26A...411L...1W} {411, L1}

\makeatother
\end{thebibliography}
\label{lastpage}
\clearpage

\setcounter{append}{0}

\onecolumn
\begin{landscape}

\begin{center}
\begin{minipage}{1.15\textwidth}
  {\bf Appendix~1.} The complete catalog of hard X-ray sources detected in the
  combined survey of three fields: M81, LMC and 3C 273/Coma. The
  descripton of the columns can be found in Sect.~2.3 of the paper.
\end{minipage}

\begin{longtable}{|r|l|r|r|r|r@{$\pm$}l|c|c|c|c|l|} 

\hline
Id&Name$^1$&R.A.&Dec.&S/N&\multicolumn{2}{c}{Flux} & D &z&$\log{L}$&Type&Notes\\
   & &deg&deg &        & \multicolumn{2}{c}{mCrab} &Mpc& & erg s$^{-1}$  & &\\ \hline
\endfirsthead

\multicolumn{12}{c}%
{{\bfseries \tablename\ \thetable{} -- continued from previous page}} \\
\hline 
Id&Name$^1$&R.A.&Dec.&S/N&\multicolumn{2}{c}{Flux} & D &z&$\log{L}$&Type&Notes\\
   & &deg&deg &        & \multicolumn{2}{c}{mCrab} &Mpc& & erg s$^{-1}$  & &\\ \hline
\endhead

\hline \multicolumn{12}{|r|}{{Continued on next page}} \\ \hline
\endfoot

\hline
\endlastfoot
\multicolumn{12}{|c|}{}\\
\multicolumn{12}{|c|}{{\bf M81 field}}\\* 
\multicolumn{12}{|c|}{}\\
\hline
\label{append:fullcat}
1&Mrk 3&93.950&71.039&39.4&5.81&0.15 &  &0.013&43.56&Sy2&\\
2&IGR J06253+7334&96.370&73.585&7.6&0.99&0.13 & &&&CV&MU Cam\\
3&Mrk 6&103.043&74.427&22.4&2.37&0.11 & &0.019&43.46&Sy2&\\
4&{\bf IGR J06571+7802} &104.277&78.044&4.2&0.47&0.11 & & & & &\\
5&QSO B0716+714&110.576&71.304&5.9&0.50&0.08 & &0.300&45.34&Blazar&\\
6&{\bf IGR J07563+5919} &119.091&59.321&4.0&0.62&0.16 & & & & & \\
7&PG 0804+761&122.929&76.034&9.4&0.63&0.07 & &0.100&44.39&Sy1&\\
8&QSO B0836+710&130.333&70.905&62.2&3.27&0.05 & &2.172&48.24&Blazar&\\
9&{\bf IGR J08501+6630} &132.547&66.515&6.9&0.38&0.06 & & & & &  \\
10&IGR J08557+6420&133.839&64.391&11.2&0.67&0.06 & &0.036&43.49&Sy2&MCG +11-11-032\\
11&NGC 2655&133.901&78.250&6.3&0.43&0.07 & 24.4& &41.64&LINER&\\
12&Mrk 110&141.292&52.292&12.2&3.44&0.28 & &0.035&44.19&Sy1&\\
13&IGR J09253+6929&141.455&69.481&13.1&0.61&0.05 & &0.039&43.52&Sy1.5&2MASX J09254750+6927532\\
14&SWIFT J0929.7+6232&142.413&62.556&10.2&0.62&0.06 & &0.026&43.15&Sy2&2MASX J09293791+6232382\\
15&SWIFT J0935.9+6120&143.991&61.314&4.5&0.29&0.07 & &0.039&43.20&Sy1&MCG +10-14-025 \\
16&SWIFT J0950.5+7318&147.509&73.248&9.3&0.45&0.05 & &0.058&43.74&Sy2/NLRG&VII Zw 292 \\
17&M81&148.898&69.080&19.8&0.90&0.05 & 3.7& &40.33&LINER&\\
18&M82 X--1&148.973&69.675&7.5&0.34&0.05 & & & &ULX&\\
19&SWIFT J1001.7+5543&150.489&55.709&11.7&1.55&0.13 & 19.1& &41.99&Sy2&NGC 3079 \\
20&{\bf IGR J10252+6716}&156.324&67.273&7.9&0.38&0.05 & &0.039&43.31&Sy2&2RXP J102510.2+671801,\\
&&&&&\multicolumn{2}{c|}{}&&&&       &KUG 1021+675 \\ 
21&SWIFT J1033.6+7303&158.576&73.016&8.3&0.41&0.05 & &0.022&42.84&XBONG&CGCG 333-038 \\
22&{\bf IGR J10380+8435} &159.513&84.587&4.4&0.63&0.14 & & & & & \\
23&SWIFT J1044.1+7024&161.072&70.431&10.2&0.51&0.05 & &0.034&43.3&Sy2&MCG +12-10-067 \\
24&QSO J1044+8054   &161.240&80.862&5.7&0.50&0.09 & &1.260&46.84&Blazar&\\
25&{\bf IGR J11015+7224}&165.392&72.409&4.0&0.22&0.05 & &1.459&46.64&Blazar& 1SXPS J110148.6+722534,\\
&&&&&\multicolumn{2}{c|}{}&&&&       & 4C 72.16\\
26&{\bf IGR J11030+7027} &165.758&70.463&4.1&0.22&0.05 & & & & &1RXS J110257.6+702948 \\
27&SWIFT J1105.7+5854&166.456&58.913&6.0&0.67&0.11 & &&&&Double source (Sy2+blazar)\\
28&NGC 3516&166.694&72.566&31.9&1.76&0.06 & 52.5& &42.92&Sy1.5&\\
29&{\bf IGR J11079+7106}&166.992&71.109&6.6&0.36&0.05 & &0.060&43.68&AGN& 1RXS J110748.8+710538,\\
&&&&&\multicolumn{2}{c|}{}&&&&       &2MASX J11074777+7105326  \\
30&SWIFT J1114.3+7944&168.978&79.698&5.0&0.42&0.08 & &0.037&43.31&Sy2&MCG +13-08-056 \\
31&SWIFT J1136.7+6738&174.155&67.595&6.8&0.47&0.07 & &0.134&44.53&Blazar&2MASX J11363009+6737042 \\
32&SWIFT J1142.7+7149&175.909&71.686&6.2&0.42&0.07 & & & &CV&DO Dra\\
33&SWIFT J1143.7+7942&176.199&79.681&9.1&0.85&0.09 & 27.2& &42.03&Sy1.2&UGC 06728 \\
34&{\bf IGR J12171+7047}&184.288&70.797&5.2&0.48&0.09 &25.0& &41.71&AGN&1SXPS J121726.3+704806,\\
&&&&&\multicolumn{2}{c|}{}&&&&       & NGC 4250  \\
35&Mrk 205&185.440&75.305&8.2&0.77&0.09 & &0.071&44.16&Sy1&\\
36&3PBC J1231.3+7044&187.944&70.746&4.6&0.49&0.11 & &0.208&44.97&Sy1.2&2MASS J12313656+7044144 \\
37&{\bf IGR J12418+7805} &190.462&78.084&4.2&0.49&0.12 & &0.022&42.91&Sy1.9& RX J1242.8+7807,   \\ 
&&&&&\multicolumn{2}{c|}{}&&&&       &NPM1G +78.0048 \\
\hline
\multicolumn{12}{|c|}{}\\
\multicolumn{12}{|c|}{{\bf LMC field}}\\
\multicolumn{12}{|c|}{}\\
\hline
38&IGR J01054--7253&16.223&-72.886&12.4&1.24&0.10 & & & &HMXB&2MASS J01044227-7254036\\
39&SMC X--1&19.285&-73.441&179.0&17.72&0.10 & & & &HMXB&\\
40&SWIFT J0157.8--7300&29.284&-73.063&5.2&0.52&0.10 & & & &HMXB&USNO-B1.0 0170-00064697 \\
41&SWIFT J0208.4--7428&31.675&-74.470&4.0&0.40&0.10 & & & &HMXB&\\
42&SWIFT J0308.5--7251&46.950&-72.819&4.6&0.43&0.09 & &0.028&43.07&Sy1.2&ESO 031-8 \\
43&IGR J03532--6829&58.344&-68.556&6.1&0.51&0.08 & &0.087&44.17&Blazar&PKS 0352-686\\
44&{\bf IGR J03574--6602} &59.375&-66.043&4.0&0.37&0.09 & &&&& \\
45&SWIFT J0422.7--5611&65.542&-56.183&4.5&0.91&0.20 & &0.043&43.78&Sy2&ESO 157-23 \\
46&1H 0419--577&66.533&-57.184&7.6&1.27&0.17 & &0.104&44.73&Sy1.5&2MASS J04260071-5712017\\
47&IGR J04288--6702&67.194&-67.075&6.0&0.43&0.07 & &&&&  \\
48&Abell 3266&67.869&-61.462&6.6&0.64&0.10 & &0.059&43.91&Cluster&\\
49&{\bf IGR J04379--7240} &69.492&-72.669&4.3&0.29&0.07 & &&&& \\
50&SWIFT J0440.2--5941&70.056&-59.662&4.1&0.44&0.11 & &0.058&43.73&Sy2&ESO 118-033 \\
51&SWIFT J0451.5--6949&72.811&-69.795&18.2&1.14&0.06 & &&&HMXB&\\
52&ESO 033--G002&74.014&-75.526&12.7&0.91&0.07 & &0.018&43.03&Sy2&\\
53&IGR J05007--7047&75.235&-70.739&13.9&0.85&0.06 & &&&HMXB&\\
54&SWIFT J0504.6--7345&76.108&-73.810&7.3&0.48&0.07 & &0.045&43.54&Sy1.9&2MASX J05043414-7349269 \\
55&SWIFT J0505.6--6735&76.365&-67.567&10.0&0.62&0.06 & &&&AGN&2MASX J05052442-6734358 \\
56&{\bf IGR J05104--6910} &77.622&-69.170&4.3&0.26&0.06 & &&&& \\
57&RX J0520.5--6932&80.060&-69.480&9.1&0.54&0.06 & &&&HMXB&[HP99] 946\\
58&LMC X--2&80.225&-71.939&10.9&0.66&0.06 & &&&LMXB&\\
59&IGR J05305--6559*&82.864&-65.935&11.2&0.70&0.06 & &&&HMXB&\\
60&LMC X--4&83.197&-66.371&334.7&20.33&0.06 & &&&HMXB&2MASS J05324953-6622132 \\
61&{\bf IGR J05329--7051} &83.226&-70.853&4.2&0.25&0.06 & &1.238&46.52&Blazar?&3XMM J053257.8-705112,\\
&&&&&\multicolumn{2}{c|}{}&&&&       &MQS J053258.11-705112.9\\
62&IGR J05346--5759&83.674&-57.992&7.9&0.87&0.11 & &&&CV&TW Pic \\
63&{\bf IGR J05347--6015} &83.678&-60.258&10.0&0.86&0.09 & &0.057&44.01&Sy1& 1SXPS J053430.8-601617,\\
&&&&&\multicolumn{2}{c|}{}&&&&       &2MASX J05343093-6016153 \\
64&PSR J0537--6910 &84.008&-69.124&8.2&0.31&0.06 & &&&Pulsar&\\
65&{\bf IGR J05373--8424} &84.342&-84.408&4.5&0.72&0.16 & &&&& \\
66&LMC X--3&84.714&-64.069&10.7&0.70&0.07 & &&&HMXB&\\
67&LMC X--1&84.856&-69.759&26.1&1.53&0.06 & &&&HMXB&2MASS J05393883-6944356 \\
68&PSR B0540--69.3&85.017&-69.327&34.4&2.01&0.06 & &&&Pulsar&\\
69&IGR J05414--6858*&85.387&-68.944&12.0&0.71&0.06 & &&&HMXB&\\
70&SWIFT J0541.5--6826&85.420&-68.398&6.2&0.36&0.06 & &&&HMXB&XMMU J054134.7-682550 \\
71&{\bf IGR J06075--6148} &91.899&-61.814&4.2&0.34&0.08 & 19.9&&41.36&AGN&3XMM J060730.3-614827,\\
&&&&&\multicolumn{2}{c|}{}&&&&       &ESO 121--G006 \\
72&SWIFT J0623.3--6438&95.784&-64.580&6.0&0.43&0.07 & &0.129&44.46&Blazar&2MASX J06230765-6436211 \\
73&IGR J06239--6052&95.925&-60.987&14.8&1.36&0.09 & &0.040&43.90&Sy2&ESO 121-28\\
74&SWIFT J0634.7--7445&98.712&-74.764&5.8&0.43&0.07 & &0.112&44.33&Sy1&2MASS J06340353-7446377\\
75&IGR J06354--7516&98.892&-75.249&9.8&0.74&0.08 & &0.653&46.31&Blazar&PKS 0637-752\\
76&{\bf IGR J06380--7536} &99.525&-75.616&4.1&0.31&0.08 & &0.089&43.97&Sy1.8&2E 0639.5-7535,\\
&&&&&\multicolumn{2}{c|}{}&&&&       &2MASX J06374318-7538458 \\
77&{\bf IGR J06503--7742} &102.599&-77.701&4.4&0.41&0.09 & &0.0373&43.30&AGN&XMMSL1 J064954.6-774216,\\
&&&&&\multicolumn{2}{c|}{}&&&&       &2MASX J06495436-7742143 \\
78&{\bf IGR J06569--6534} &104.240&-65.570&6.2&0.51&0.08 & &0.0305&43.22&Sy1&  RX J065630-65349,\\
&&&&&\multicolumn{2}{c|}{}&&&&       &Fairall 265  \\
79&{\bf IGR J07296--5854} &112.413&-58.905&4.2&0.72&0.17 & &&&&\\
80&SWIFT J0747.6--7326&116.989&-73.449&6.8&0.70&0.10 & &0.036&43.51&LINER?&2MASX J07473839-7325533 \\
81&EXO 0748--676&117.097&-67.756&28.7&3.25&0.11 & &&&LMXB&\\
82&SWIFT J0826.2--7033&126.584&-70.527&9.2&1.31&0.14 & &&&&1SXPS J082623.1-703143 \\
83&IGR J09025--6814&135.680&-68.219&4.3&0.68&0.15 & &0.013&42.62&XBONG&NGC 2788A\\
\hline
\multicolumn{12}{|c|}{}\\
\multicolumn{12}{|c|}{{\bf 3C 273/Coma field}}\\
\multicolumn{12}{|c|}{}\\
\hline
84&SWIFT J1144.1+3652&176.118&36.924&5.9&0.56&0.10 & &0.038&43.46&Sy1&KUG 1141+371 \\
85&{\bf IGR J11477+0557} &176.931&5.966&4.6&0.40&0.09 & &&&& \\
86&SWIFT J1148.3+0901&177.025&9.049&4.2&0.35&0.08 & &0.069&43.79&Sy1.5&2MASX J11475508+0902284 \\
87&SWIFT J1148.7+2941&177.171&29.609&5.1&0.72&0.14 & &0.023&43.12&Sy1&MCG +05-28-032 \\
88&3PBC J1152.9+3307&178.186&33.104&4.0&0.43&0.11 & &1.398&46.89&Blazar&7C 1150+3324\\
89&SWIFT J1200.8+0650&180.237&6.810&11.9&0.75&0.06 & &0.036&43.54&Sy2&2MASX J12005792+0648226\\
90&SWIFT J1201.2--0341&180.334&-3.696&5.6&0.58&0.10 & &0.020&42.88&Sy1&Mrk 1310 \\
91&{\bf IGR J12024--1127} &180.622&-11.460&4.3&0.91&0.21 & &&&& \\
92&NGC 4051&180.769&44.522&19.9&2.18&0.11 & 14.0&&41.86&Sy1&NGC 4051\\
93&{\bf IGR J12038--1210} &180.958&-12.178&4.0&0.87&0.22 & &&&& \\
94&NGC 4074&181.115&20.324&8.8&0.94&0.11 & &0.022&43.21&Sy2&NGC 4074\\
95&3PBC J1204.7+3109&181.121&31.193&4.9&0.55&0.11 & &0.025&43.08&Sy1.9&UGC 7064 \\
96&SWIFT J1207.5+3355&181.949&33.854&4.6&0.44&0.10 & &0.079&44.01&Sy2&B2 1204+34 \\ 
97&SWIFT J1209.5+4702&182.323&47.036&7.3&0.98&0.13 & &0.024&43.30&Sy2&Mrk 198 \\
98&NGC 4138&182.363&43.688&12.8&1.33&0.10 & 20.7&&41.99&Sy1.9&\\
99&{\bf IGR J12095--0420} &182.394&-4.344&4.0&0.38&0.09 & &&&& \\
100&NGC 4151&182.628&39.408&305.2&27.67&0.09 & 13.4&&42.93&Sy1.5&NGC 4151\\
101&IGR J12107+3822&182.671&38.343&10.8&0.97&0.09 & &0.023&43.24&Sy1.5&KUG 1208+386\\
102&NGC 4180&183.253&7.036&10.7&0.57&0.05 & 39.2& &41.97&LINER&\\
103&Was 49&183.571&29.578&5.2&0.56&0.11 & &0.061&43.88&Sy1&\\
104&IGR J12172+0710&184.302&7.187&45.3&2.32&0.05 & 25.0&&42.39&Sy1.2&NGC 4235 \\
105&Mrk 766&184.600&29.828&13.2&1.35&0.10 & &0.013&42.89&Sy1&\\
106&NGC 4258&184.727&47.298&7.1&0.99&0.14 & 7.5&&40.97&Sy2&\\
107&3PBC J1220.1+0203&185.022&2.065&4.1&0.23&0.06 & &0.240&44.78&Sy1.8&  PKS 1217+023\\
108&{\bf IGR J12204+0452} &185.120&4.868&4.6&0.24&0.05 & &&&&\\
109&{\bf IGR J12208--0711} &185.212&-7.192&4.5&0.46&0.10 & &&&&\\
110&QSO B1218+304&185.351&30.169&6.0&0.60&0.10 & &0.184&44.94&Blazar&\\
111&{\bf IGR J12224+0306} &185.606&3.110&4.3&0.23&0.05 & &0.255&44.84&QSO?&1SXPS J122208.8+030717,\\
&&&&&\multicolumn{2}{c|}{}&&&&       &SDSS J122208.78+030718.4\\
112&4C 04.42&185.620&4.239&16.3&0.84&0.05 & &0.966&46.78&Blazar&\\
113&Mrk 50&185.851&2.689&16.0&0.86&0.05 & &0.023&43.21&Sy1.2&\\
114&QSO B1222+216&186.234&21.363&9.5&0.83&0.09 & &0.433&45.93&Blazar&\\
115&NGC 4388&186.449&12.663&155.7&8.36&0.05 & 20.6&&42.78&Sy2&\\
116&NGC 4395&186.456&33.552&13.6&1.22&0.09 & 4.5&&40.62&Sy1.8&\\
117&3C 273&187.279&2.052&249.7&13.46&0.05 & &0.158&46.15&Blazar&\\
118&{\bf IGR J12304+0946} &187.620&9.776&4.2&0.21&0.05 & &&&&\\
119&{\bf IGR J12375+2156} &189.392&21.944&4.3&0.36&0.08 & &&&&\\
120&NGC 4579&189.412&11.804&9.5&0.50&0.05 & 19.8&&41.52&LINER?&\\
121&SWIFT J1238.6+0928&189.665&9.440&9.2&0.46&0.05 & &0.032&43.22&AGN&VCC 1759 \\
122&NGC 4593&189.925&-5.356&47.3&3.86&0.08 & 44.0&&43.10&Sy1&\\
123&SWIFT J1240.9+2735 &190.260&27.505&5.4&0.50&0.09 & &0.057&43.77&Sy2&KUG 1238+278A\\
124&{\bf IGR J12412+3007} &190.302&30.125&5.4&0.49&0.09	 & &&&&  \\
125&NGC 4736&192.686&41.151&4.5&0.50&0.11 & 5.0&&40.33&LINER&\\
126&NGC 4748&193.078&-13.415&5.9&0.95&0.16 & &0.015&42.84&NLS1&\\
127&{\bf IGR J12546+1139} &193.672&11.663&4.6&0.27&0.06 & &0.873&46.18&Blazar&RX J1254.6+1141,\\
&&&&&\multicolumn{2}{c|}{}&&&&       &QSO B1252+119 \\
128&3C 279&194.025&-5.806&11.9&1.10&0.09 & &0.536&46.28&Blazar&\\
129&Coma Cluster&194.892&27.932&13.0&1.18&0.09 & &0.023&43.34&Cluster&\\
130&SWIFT J1300.1+1635&195.072&16.545&4.7&0.35&0.08 & &0.080&43.93&Sy1&2MASX J13000533+1632151 \\
131&Mrk 783&195.741&16.396&12.7&0.97&0.08 & &0.067&44.21&NLS1&\\
132&NGC 4941&196.042&-5.569&6.9&0.70&0.10 & 21.2&&41.73&Sy2&\\
133&NGC 4939&196.067&-10.347&8.3&1.15&0.14 & 38.8&0.010&42.62&Sy2&\\
134&IGR J13091+1137&197.288&11.641&27.1&1.98&0.07 & &0.025&43.64&XBONG&NGC 4992\\
135&{\bf IGR J13100+0830} &197.507&8.508&4.1&0.29&0.07 & &&&&\\
136&IGR J13133--1109&198.319&-11.177&4.1&0.68&0.17 & &0.034&43.45&Sy1&2MASX J13130580-1107424\\
137&NGC 5033&198.357&36.583&5.2&0.64&0.12 & 18.7&&41.58&Sy1.9&\\
138&{\bf IGR J13142--0500} &198.557&-5.014&4.1&0.47&0.12 & &&&&\\
139&IGR J13149+4422&198.823&44.434&8.2&1.14&0.14 & &0.037&43.73&Sy2& Mrk 248 \\
140&{\bf IGR J13169+3733} &199.237&37.553&4.1&0.54&0.13 & &0.195&44.95&AGN&RX J1317.0+3735,\\
&&&&&\multicolumn{2}{c|}{}&&&&       &2MASS J13170290+3735329	 \\
141&SWIFT J1321.2+0859&200.271&8.941&7.3&0.63&0.09 & &0.032&43.35&LINER?&NGC 5100 NED02\\
142&{\bf IGR J13310--1355} &202.756&-13.931&4.0&1.18&0.29 & &&&&\\
143&NGC 5252&204.555&4.550&43.5&4.94&0.11 & &0.023&43.96&Sy1.9&\\
144&Mrk 268&205.300&30.392&8.8&1.18&0.13 & &0.040&43.82&Sy2&\\
145&3PBC J1342.0+3539&205.546&35.699&5.9&1.01&0.17 & 16.0&&41.64&Sy1.9&NGC 5273\\
146&IGR J13466+1921&206.696&19.404&4.6&0.57&0.12 & &0.084&44.18&Sy1.2&2MASX J13462846+1922432\\
147&{\bf IGR J13486+1554} &207.168&15.901&4.8&0.57&0.12 & &&&&\\
\end{longtable}

\begin{minipage}{1.15\textwidth}
$^{1}$ The names of the sources previously unknown in hard X-ray band ($17-60$~keV) are highlighted in bold. The sources in spatial confusion are indicated by star. The measured flux of the sources in spatial confusion should be taken with the caution.\\
\end{minipage}
\end{center}	
\end{landscape}
\twocolumn
\end{document}